\documentclass{aa}
\usepackage{epsfig}
\usepackage{natbib}
\usepackage{graphicx}
\begin{document}
\newcommand{\bea}{\begin{eqnarray}}    
\newcommand{\eea}{\end{eqnarray}}      
\newcommand{\be}{\begin{equation}}
\newcommand{\ee}{\end{equation}}
\newcommand{\bef}{\begin{figue}}
\newcommand{\eef}{\end{figure}}
\newcommand{\etal}{et al.}
\newcommand{\kms}{\,{\rm km}\;{\rm s}^{-1}}
\newcommand{\hubunits}{\,\kms\;{\rm Mpc}^{-1}}
\newcommand{\hmpc}{\,h^{-1}\;{\rm Mpc}}
\newcommand{\hkpc}{\,h^{-1}\;{\rm kpc}}
\newcommand{\msun}{M_\odot}
\newcommand{\K}{\,{\rm K}}
\newcommand{\cm}{{\rm cm}}
\newcommand{\cd}{{\langle n(r) \rangle_p}}
\newcommand{\Mpc}{{\rm Mpc}}
\newcommand{\kpc}{{\rm kpc}}
\newcommand{\xir}{{\xi(r)}}
\newcommand{\xrp}{{\xi(r_p,\pi)}}
\newcommand{\xsirpi}{{\xi(r_p,\pi)}}
\newcommand{\wrp}{{w_p(r_p)}}
\newcommand{\gr}{{g-r}}
\newcommand{\Navg}{N_{\rm avg}}
\newcommand{\Mmin}{M_{\rm min}}
\newcommand{\fiso}{f_{\rm iso}}
\newcommand{\Mr}{M_r}
\newcommand{\rp}{r_p}
\newcommand{\zmax}{z_{\rm max}}
\newcommand{\zmin}{z_{\rm min}}

\def\eg{{e.g.}}
\def\ie{{i.e.}}
\def\spose#1{\hbox to 0pt{#1\hss}}
\def\ltapprox{\mathrel{\spose{\lower 3pt\hbox{$\mathchar"218$}}
\raise 2.0pt\hbox{$\mathchar"10C$}}}
\def\gtapprox{\mathrel{\spose{\lower 3pt\hbox{$\mathchar"218$}}
\raise 2.0pt\hbox{$\mathchar"10E$}}}
\def\inapprox{\mathrel{\spose{\lower 3pt\hbox{$\mathchar"218$}}
\raise 2.0pt\hbox{$\mathchar"232$}}}

\title{Breaking the self-averaging properties of spatial galaxy
  fluctuations in the Sloan Digital Sky Survey - Data Release Six}

\subtitle{}

\author{Francesco Sylos Labini \inst{1,2}, Nikolay L. Vasilyev \inst{3}, Yurij 
V. Baryshev \inst{3}}

\titlerunning{Breaking of self-averaging in SDSS-DR6}
\authorrunning{Sylos Labini, Vasilyev, Baryshev}

\institute{ 
Centro Studi e Ricerche Enrico Fermi, Via Panisperna 89 A, 
Compendio del Viminale, 00184 Rome, Italy
\and Istituto dei Sistemi Complessi CNR, 
Via dei Taurini 19, 00185 Rome, Italy.
\and 
Institute of Astronomy, St.Petersburg 
State University, Staryj Peterhoff, 198504,
St.Petersburg, Russia
}

\date{Received / Accepted}

\abstract{ Statistical analyses of finite sample distributions usually
  assume that fluctuations are self-averaging, i.e.  statistically
  similar in different regions of the given sample volume. By using
  the scale-length method, we test whether this assumption is
  satisfied in several samples of the Sloan Digital Sky Survey Data
  Release Six. We find that the probability density function (PDF) of
  conditional fluctuations, if filtered on large enough spatial scales
  (i.e., $r>30$ Mpc/h), shows relevant systematic variations in
  different subvolumes of the survey. Instead for scales of $r<30$
  Mpc/h, the PDF is statistically stable, and its first moment
  presents scaling behavior with a negative exponent around one. Thus
  while up to 30 Mpc/h galaxy structures have well-defined power-law
  correlations, on larger scales it is not possible to consider whole
  sample average quantities as meaningful and useful statistical
  descriptors. This situation stems from galaxy structures
  corresponding to density fluctuations that are too large in
  amplitude and too extended in space to be self-averaging on such
  large scales inside the sample volumes: galaxy distribution is
  inhomogeneous up to the largest scales, i.e. $r \approx 100$ Mpc/h
  probed by the SDSS samples.  We show that cosmological corrections,
  such as K-corrections and standard evolutionary corrections, do not
  qualitatively change the relevant behaviors. We consider in detail
  the relation between several statistical measurements generally used
  to quantify galaxy fluctuations and the scale-length analysis by
  discussing how the breaking of self-averaging properties makes it
  impossible a reliable estimation of average fluctuations amplitude,
  variance, and correlations for $r>30$ Mpc/h.  Finally we show that
  the large-amplitude galaxy fluctuations observed in the SDSS samples
  are at odds with the predictions of the standard $\Lambda$CDM model
  of structure formation.
 } 
  
\maketitle

\section{Introduction}

The statistical characterization of galaxy structures represents a
central problem for our { understanding} of the large-scale universe.
Once three-dimensional galaxy samples are provided by observations,
one may think the problem is relatively simple; i.e., all that remains
to do is to characterize the statistical properties of $N$ points
(galaxies) contained in a volume $V$. However, there are several
issues that must be considered with great care; namely, (i) the
definition of the statistical methods employed and analysis of the
assumptions implicitly used by them; (ii) construction of the samples
and the consideration of cosmological corrections; (iii) comparison of
results in galaxy catalogs with model predictions.  Even though each
of these issues requires a separate discussion, sometimes in the
literature the reliability of statistical methods is hidden by the
problems related to cosmological corrections and/or by sampling (or
biasing) of a given distribution, which is the problem to be considered
when comparing results of observations with theoretical predictions
and cosmological N-body simulations
\footnote{ In general particles in cosmological N-body simulations are
  supposed to represent a coarse grained distribution of the
  microscopic dark matter particles. From the N-body dark matter
  particles one constructs the galaxy density field, by using certain
  procedures, which can be generally thought of as a sampling
  mechanism. The key element of this selection is that galaxies are
  supposed to form on the highest density peaks of the underlying dark
  matter field.}.
In this way one is not able to properly disentangle the different
problems and to ask the relevant questions at each step. For this
reason, in what follows we try to discuss the three issues above by
considering each in turn. In particular, only when there is agreement
about the statistical methods used will it be possible to compare
clearly results from different authors and to isolate the problems
related to cosmological corrections and/or sampling.


There has been an intense debate about the most suitable statistical
methods for characterizing galaxy properties, particularly galaxy
correlations \citep[see][]
{pie87,davis,pmsl96,slmp98,jmsl,wu,gsl01,hogg,joyce05,bt05,
  2df_paper,dr4_paper,paper_2df_prl,paper_2df_aea,paper_sdss1}.  { The
  most suitable statistical method for characterizing the properties
  of a given stochastic point process depends on the underlying
  correlations of the point distribution itself.  There can be
  different situations for the statistical properties of any set of
  points (in the present case, galaxies) in a finite sample.  Let us
  briefly consider four different cases.  {\it Inside a given sample}
  the galaxy distribution is approximated by a {\it uniform}
  stochastic point process, or in other words, {\it inside a given
    sample} the average density is well-defined. This means that the
  density, measured for instance in a sphere of radius $r$ randomly
  placed inside the sample, has small fluctuations. In this situation
  the relative fluctuations between the average density estimator and
  the ``true'' density is less than unity. Density fluctuations
  may be correlated, and the correlation function can be {\it (i)}
  short-ranged (e.g., exponential decay) or {\it (ii)} long-ranged
  (e.g., power law). In other words these two cases correspond to a
  uniform stochastic point process with (i) short-range and (ii)
  long-range correlations.

On the other hand, it may happen that galaxy distribution is not
uniform. In this situation, the density measured, for instance, in a
sphere of radius $r$ randomly placed inside the sample, has large
fluctuations; i.e., it varies wildly in different regions of the
sample.  In this situation the point distribution can generally
present long-range correlations of {\it large amplitude}. Then it may,
case {\it (iii)}, or may not, case {\it (iv)}, present self-averaging
properties, depending on whether  measurements of the density in
different subregions show systematic (i.e., not statistical)
differences that depend, for instance, on the spatial positions of the
specific subregions.  When this is so, the considered statistics are
not statistically self-averaging in space because the PDF
systematically differs in different subregions and whole-sample
average values are not meaningful descriptors. In general, such
systematic differences may be related to two different possibilities:
(i) the underlying distribution is not translationally and/or
rotationally invariant, (ii) the volumes considered are not large
enough for fluctuations to be self-averaging.

In determining statistical properties, a fundamental assumption is
very often used in the finite-sample analysis: that sample density is
supposed to provide a reliable estimate of the ``true'' space density,
i.e., that the point distribution is well-represented by cases (i) or
(ii) above.  This corresponds to the {\it assumption} the relative
fluctuations between the average density estimator and the ``true''
density are smaller than unity.  In general, this is a very strong
assumption that may lead to underestimating finite size effects in the
statistical analysis.

For instance, let us suppose that the distribution {\it inside the
  given sample} is {\it not} uniform, i.e. cases (iii) and (iv) above.
In this case the results of the statistical analysis are biased by
important finite-size effects, so that all estimations of statistical
quantities based on the uniformity assumption (i.e. the two-point
correlation function and all quantities normalized to the sample
average) are affected, on all scales, by this {\it a-priori}
assumption that is inconsistent with the data properties \citep{book}.
In addition, while for case (iii) one may consider a class of whole
sample-averaged quantities, i.e. conditional
statistics~\footnote{Conditional statistics are not normalized to the
  sample density estimation (see Sect.2).} in case (iv) these become
meaningless.

For this reason, our first aim is to study whether galaxy distribution
is self-averaging by characterizing conditional fluctuations. If the
distribution is self-averaging, then one can consider a whole-sample
average quantity and study the possible transition from non-uniformity
to uniformity by characterizing the behavior of, for instance, the
conditional density. If the distribution is uniform, or becomes
uniform on a certain scale smaller than the sample size, one can
characterize the (residual) correlations between density fluctuations
by studying the standard two-point correlation function. Therefore the
consideration of $\xi(r)$ is the last point on this list, and it is
appropriate only if one has proved that the distribution is
self-averaging and uniform inside the given sample.  }

These issues are relevant in studies of the galaxy distribution
because in the past twenty years it is has been observed that galaxy
structures are organized in a complex network of clusters, filaments,
and voids on scales up to hundreds of Mpc
\citep[see][]{Kirshner,gh89,brod,pp,gott,einasto1,einasto2}.  From the
statistical point of view, the problem is whether these structures are
compatible with the very small characteristic length scale of the
galaxy distribution of about ten Mpc. This is the scale at which the
two-point correlation function is equal to unity, and it has been
measured to be in the range of 5-15 Mpc/h in different (angular and
three-dimensional) catalogs
\citep{tk69,dp83,davis88,park,benoist,norbergxi01,
  norbergxi02,zehavi_earlydata,zehavietal05}.  The essence of the
problem is not whether these measurements have been properly made as
indeed they have been, but other whether the statistical methods used
to get this result are consistent with the properties of the galaxy
distribution in these samples \citep[see][]{book,paper_sdss1}.


By measuring the redshift-dependent luminosity function and the
comoving radial density of galaxies in the Sloan Digital Sky Survey
(SDSS) Data Release 1 (DR1), it has been found that the apparent
number density of bright galaxies increases by a factor $\approx $ 3
as redshift increases from $z = 0$ to $z = 0.3$ \citep{loveday}.  To
explain these observations, a significant evolution in the luminosity
and/or number density of galaxies at redshifts $z < 0.3$ has then been
proposed \citep{loveday}.  However, an independent test has not been
provided to support such a conclusion; in particular, the possible
effect of large density fluctuations on the basic assumptions used in
this analysis (i.e. large-scale uniformity of the density field) was
not tested, although it was noticed that these results do not preclude
significant density fluctuations in the local
universe on very large scales.  In what follows, we will carefully
consider these results and present a different conclusion for these
observations, namely that galaxy clustering on very large scales is
certainly making an important contribution to the observed behaviors
of galaxy counts.

Regardless of the origin of the big change in the spatial density
found by \citet{loveday}, we note that the density varies by a factor
three within the given sample implies that it is meaningless to derive
amplitudes of fluctuations with respect to this quantity.  Indeed, in
this situation the estimation of the amplitude of fluctuations
normalized to the sample density is biased by systematic effects, and
whole sample-averaged quantities, such as the two-point correlation
function and the power-spectrum, are not meaningful and stable
statistical descriptors.  Another question we address here in more
detail concerns the physical origin of the density growth. As
mentioned, while in \citet{loveday} it is concluded that the density
growth comes from evolution leaving, however, open the question of the
contribution of large scale structures, we concluded that this stems
from large-scale fluctuations \citep{paper_sdss1}. Here we show that,
if relevant on such low redshifts, galaxy evolution is not the main
cause of  the measured behaviors.  This result is reached by
performing several specific tests that include some rough
determinations of the effect of evolution as in
\citet{blanton2003,tegmark2004}.

The paper is organized as follows. In Sect.\ref{sec:methods} we give a
brief overview of our statistical methods, stressing the role of
assumptions and the properties of conditional and unconditional
fluctuations. Then in Sect.\ref{sec:samples} we discuss the procedure
used for selecting the data from the SDSS-DR6 \citep{paperdr6} archive
and the various corrections applied to constructing the samples used
in the analysis.  In Sect.\ref{sec:sl} we discuss the main results of
the statistical analysis we considered, that concerns the study of
conditional fluctuations in the SDSS samples and their PDF.  Then, in
Sect.\ref{sec:mock} we compare the conditional fluctuations in the
real galaxy samples with the predictions of theoretical models and
with those measured in mock galaxy catalogs constructed from
cosmological N-body simulations.  These are the outcome of
gravitational N-body simulations of a concordance model, i.e. a
$\Lambda$ Cold Dark Matter (CDM) model \citep{springel05}, and
represent the predictions of theoretical models for the correlation
properties of non-linear structures \citep{cronton06}.  Finally
Sect.\ref{sec:discussion} we draw our main conclusions.


\section{Overview of statistical methods} 
\label{sec:methods}

There are several {\it a priori} assumptions that are generally used
in statistical studies of galaxy samples and that require detailed
consideration \citep[see][]{book}.  Galaxy distribution is considered
to be a realization of a {\it stationary} stochastic point
process. This means that it is assumed to be statistically
translationally, and rotationally invariant, thereby satisfying the
conditions of statistical isotropy and homogeneity in order to avoid
special points or directions.  These conditions are enough to satisfy
the Copernican principle, i.e., that there are no special points or
directions; however they do not imply spatial homogeneity. Indeed an
inhomogeneous distribution can satisfy the Copernican principle even
though this is characterized by large voids and structures
\citep{sl94,pwa,book}.

\subsection{A brief summary of the statistical properties}

We now briefly discuss several properties of stochastic point
processes (SPP) that are  useful in the rest of the paper
\citep{book}.
\begin{itemize}
\item A {\it stationary } SPP (SSPP) satisfies the conditions for a 
  statistically translational and rotational invariant. It can be
  uniform (spatially homogeneous) or nonuniform (spatially
  inhomogeneous).
\item An SSPP is {\it ergodic} if the ensemble average of a
  statistical quantity characterizing its properties equals its
  infinite volume average. In a finite volume, only volume averages
  determinations are defined (i.e. estimations of statistical
  quantities). The ergodicity of an SSPP is a necessary assumption when
  one wants to compare volume average quantities with theoretical
  predictions.
\end{itemize}

Let $\rho(\vec{r})$ be a microscopic density function, that is, a
realization of a given stochastic process.  A stochastic process is
ergodic if a generic observable macroscopic variable $F =
F(\rho(\vec{r_1}),\rho(\vec{r_2}),...)$ satisfies the following
relation: the average over an ensemble of realizations $\langle F
\rangle$ is equal to the spatial average $\overline{F}$ defined by
\be 
\label{ergo}
\overline{F} = \lim_{V\rightarrow \infty} \frac{1}{V} \int_V
F(\rho(\vec{r_1}+\vec{r}),\rho(\vec{r_2}+\vec{r}),...)  d^3r \,. 
\ee 
 When $V$ in
Eq.\ref{ergo} is finite, then $\overline{F}$ is a statistical {\it
estimator} of $\langle F \rangle$ in a given sample. Therefore the
assumption of ergodicity is necessary if we want to use a statistical
estimator to verify a theoretical prediction, which is expressed in
terms of ensemble averages.

\begin{itemize}
\item An SSPP is {\it uniform} if, in a finite but large enough sample,
  fluctuations in the density are small enough. For instance, the
  scale $\lambda_0$ at which an SSPP becomes uniform can be defined to
  be scale beyond which the fluctuations on the average density
  filtered on that scale are of the same order of the average density
  itself, and then they are smaller on larger scales. To test whether
  an SSPP is uniform one can use {\it conditional} properties, which
  are defined also when the SSPP is not uniform.
\item A uniform SSPP inside a given sample has a {\it well-defined
  average} density, i.e. the sample determination is representative of
  the ensemble value within some relative small errors. Alternatively
  the amplitude of the two-point correlation function is small enough
  on large scales to guarantee that positive average density
  exists. This is, however, a necessary but not a sufficient
  condition, as the amplitude of estimator of this function can be
  small also for a non-uniform distribution in a finite sample. In the
  latter case however the amplitude is not a significant statistical
  measurement.
\item An SSPP has a well-defined {\it crossover to homogeneity}, if it
  is nonuniform on scales smaller than $\lambda_0$ and uniform on 
  larger scales $\lambda_0$. The length scale $\lambda_0$ marks the
  transition from the regime from large to small fluctuations.  At
  scales $r>\lambda_0$ one-point statistical properties
  (i.e. unconditional properties) are well defined. To study the
  approach to uniformity one should consider conditional properties.
\item A uniform SSPP can have {\it long range correlations},
  i.e. characterized by a non-zero two-point correlation function at
  all scales. This latter case describes the case of an LCDM model,
  which is indeed characterized by large scale super-homogeneity
  \citep{glass}. A system can be uniform and, at the same time,
  long-range correlated only if the amplitude of the two-point
  correlation function $\xi(r)$ is small enough on large scales.
\item The {\it range of correlations} for a uniform SSPP is measured by
  the functional behavior of the two-point correlation function
  $\xi(r)$. If the system has critical correlations, $\xi(r)$ is a
  power-law function of distance.
\item An SSPP is {\it nonuniform} (or spatially inhomogeneous), inside
  a given sample, if the conditional density does not converge to a
  constant value.  If the distribution is self-averaging (see below)
  and nonuniform then the conditional density is a varying function
  of the distance.  When this does not change anymore function of
  distance, the distribution uniform.
\item To test whether a {\it nonuniform} SSPP is {\it 
  self-averaging} in a finite volume and on a certain scale $r$, one
  may study the PDF of conditional fluctuations. If this is not
  statistically stable in different subvolumes of linear size $r$,
  then the self-averaging property is not satisfied.
\end{itemize}

 The self-averaging property is closely related to ergodicity.  In a
 volume of linear size $L$, any observable $F$ has different values
 for different realizations of the randomness (i.e. of the stochastic
 process) and is thus a stochastic variable described by a PDF
 $P(F,L)$.  By denoting the average 
\[
\overline{F}_L = \int F P(F,L)
 dF
\] 
and variance \[ 
\overline{\Delta F^2}_L = \int F^2 P(F,L) dF -
 \overline{F}_L^2 \;, \] a system is said to exhibit {\it self-averaging}
 if \citep{Aharony}
\footnote{Equivalently, if the PDF $P(F,L)$ tends to a Dirac's
    delta function for $L \rightarrow \infty$ then the system is said
    to exhibit self-averaging properties.}
\[
\lim_ {L \rightarrow \infty} \frac{\overline{\Delta
    F^2}_L}{\overline{F}_L^2} =  0 \;.
\] 

In such a case, a single large system is enough to represent the whole
ensemble.  When there are long-range correlations, the property of
self-averaging is non trivial as self-averaging requires the size $L$
of the sample to be larger than the range of correlations
\citep{Aharony}.  The concepts of ergodicity and self-averaging refer
to two different properties of a stochastic process; namely ergodicity
of the variable $F$ implies Eq.\ref{ergo} while the self-averaging
property has to be ascribed to the ensemble variable $\overline{F}_L$,
which is determined in a finite sample.

Finally it is worth noticing that, if the distribution is uniform, for
the cases in which correlations are both short or long ranged, any
global (spatially averaged) observable of the system has Gaussian-type
fluctuations, in agreement with the central limit theorem. When there
are long-range correlations of large amplitude the central limit
theorem does not hold and fluctuations in global quantities usually
have non-Gaussian fluctuations (see \citet{tibor} for a more detailed
discussion).


\subsection{A toy model}

To further clarify the concepts previously illustrated, we discuss a
simple toy model.  We generate a stochastic point distribution as
follows. We distribute randomly in two dimensional Euclidean space
rectangular sticks and the points within each stick. The center of
each stick and its orientation are chosen randomly in a box of side
$L$ (for simplicity we fix $L=1$). The points of each stick are placed
randomly within its area, which for simplicity we take to be $\ell
\times \ell/10$; these points have constant density within each stick.
The length scale $\ell$ can vary as can the number of sticks placed in
the box. Different realizations of this toy model are shown in
Fig.\ref{toy}. The conditional average density (see
Eq.\ref{first_moment} below), i.e. the average density computed in
spheres whose center is a distribution point, is shown in
Fig.\ref{fig:gamma_toy} and the PDF of the conditional density in
Fig.\ref{fig:pdf_toy}.  

By taking the  dimension $\ell$ of the sticks (in  this case equal for
all sticks) small enough and the number of sticks  large enough, one
has  generated  a  uniform  distribution with  positive  correlations,
i.e. $\xi(r)  >0$, on small  scales  (Fig.\ref{toy} upper  left panel
  --- model  T1). {  In  this case  the  average conditional  density
  (Fig.\ref{fig:gamma_toy})  rapidly decays to  a constant  value, and
  the PDF of fluctuations (Fig.\ref{fig:pdf_toy}) is approximated by a
  Gaussian  function.  When  the  dimension $\ell$  of  the sticks  is
  increased and their number still large enough, then the distribution
  is still uniform, but it  is positively correlated on larger scales.
  In the example  shown in Fig.\ref{toy} (upper right  panel --- model
  T2) the dimension  of the sticks is about the box  side, making it a
  uniform distribution  with (weak)  correlations extending up  to the
  box  size.  In  such a  situation, the  average  conditional density
  reaches  a  constant  value  on  a  scale  (the  homogeneity  scale)
  comparable            to           the            box           size
  (Fig.\ref{fig:gamma_toy}). Correspondingly, the PDF is Gaussian only
  when  fluctuations   are  filtered  on  scales   comparable  to  the
  homogeneity scale (Fig.\ref{fig:pdf_toy}).

We can then increase the dimension of the sticks further and decrease
their number (Fig.\ref{toy}, bottom left panel --- model T3). In this
case the distribution is not uniform, as there are holes as large as
the sample. The density thus presents large fluctuations and it is not
a well-defined quantity on the sample scale. This is clearly a
positive correlated distribution, with long-range correlations (up to
the sample size in this case) of large amplitude.  This is shown by
the behavior of the average conditional density
(Fig.\ref{fig:gamma_toy}), which does not converge to a constant value
inside the box.  Therefore this is not a uniform distribution; indeed,
the PDF of fluctuations (Fig.\ref{fig:pdf_toy}), filtered on large
enough spatial scales, does not converge to a Gaussian function.  To
show whether  the distribution is self-averaging inside the
simulation box, one may compare the full PDF with the ones measured in
two half parts of the box. One may see from Fig.\ref{fig:pdf_toy}
that, although there are differences, the shape of the PDF is similar
in the two subsamples. Particularly, the peak and the width of the
three PDF are approximately the same.

Finally we can take sticks with different $\ell$. In the example shown
in Fig.\ref{toy} (bottom right panel --- model T4), this is the same
for all but for a single stick that has a $\ell$ larger than the
sample size. As for the previous case this is a strongly correlated
distribution, which is not uniform inside the box.  Indeed, the
average conditional density does not flatten inside the box
(Fig.\ref{fig:gamma_toy}).  In addition, this distribution is not
self-averaging.  Indeed, by measuring the PDF of conditional
fluctuations in different regions of the sample (say the upper and the
bottom parts --- see Fig.\ref{fig:pdf_toy}), one finds systematic
(i.e., not statistical) differences. This is an effect of the strong
correlations extending well over the size of the sample.

 A quantitative measurement of the breaking of the self-averaging
 property is represented, for instance, by determining the first and
 second moment of the PDF and by checking whether they are stable in
 different subregions of the samples. One may note from
 Fig.\ref{fig:pdf_toy} that, for the model T4, both the peak and the
 width of the PDF are different when measured in different sample
 subregions or in the whole sample box, thus indicating the breaking
 of self-averaging.

\begin{figure*}
\includegraphics*[angle=0, width=0.5\textwidth]{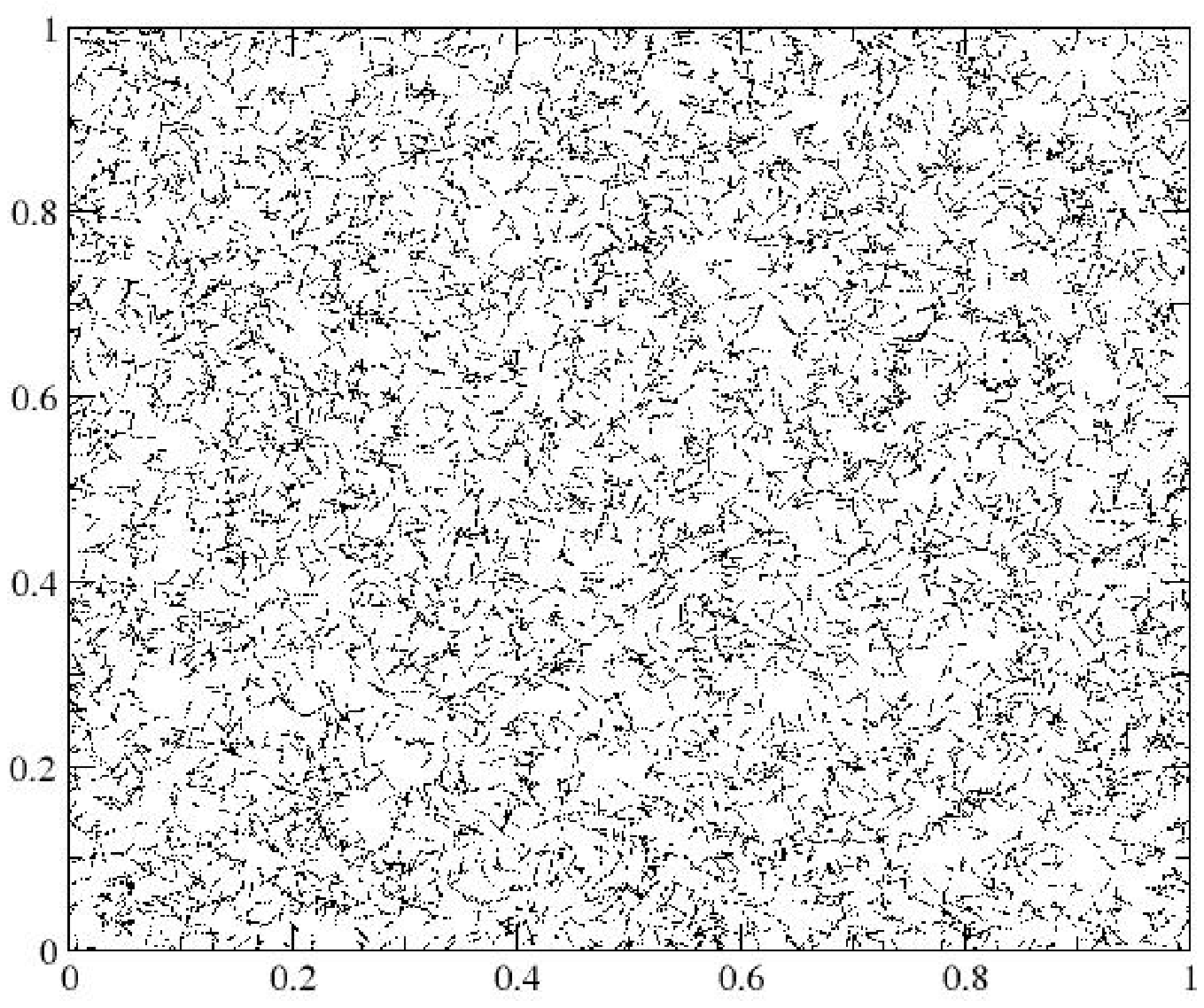}
\includegraphics*[angle=0, width=0.5\textwidth]{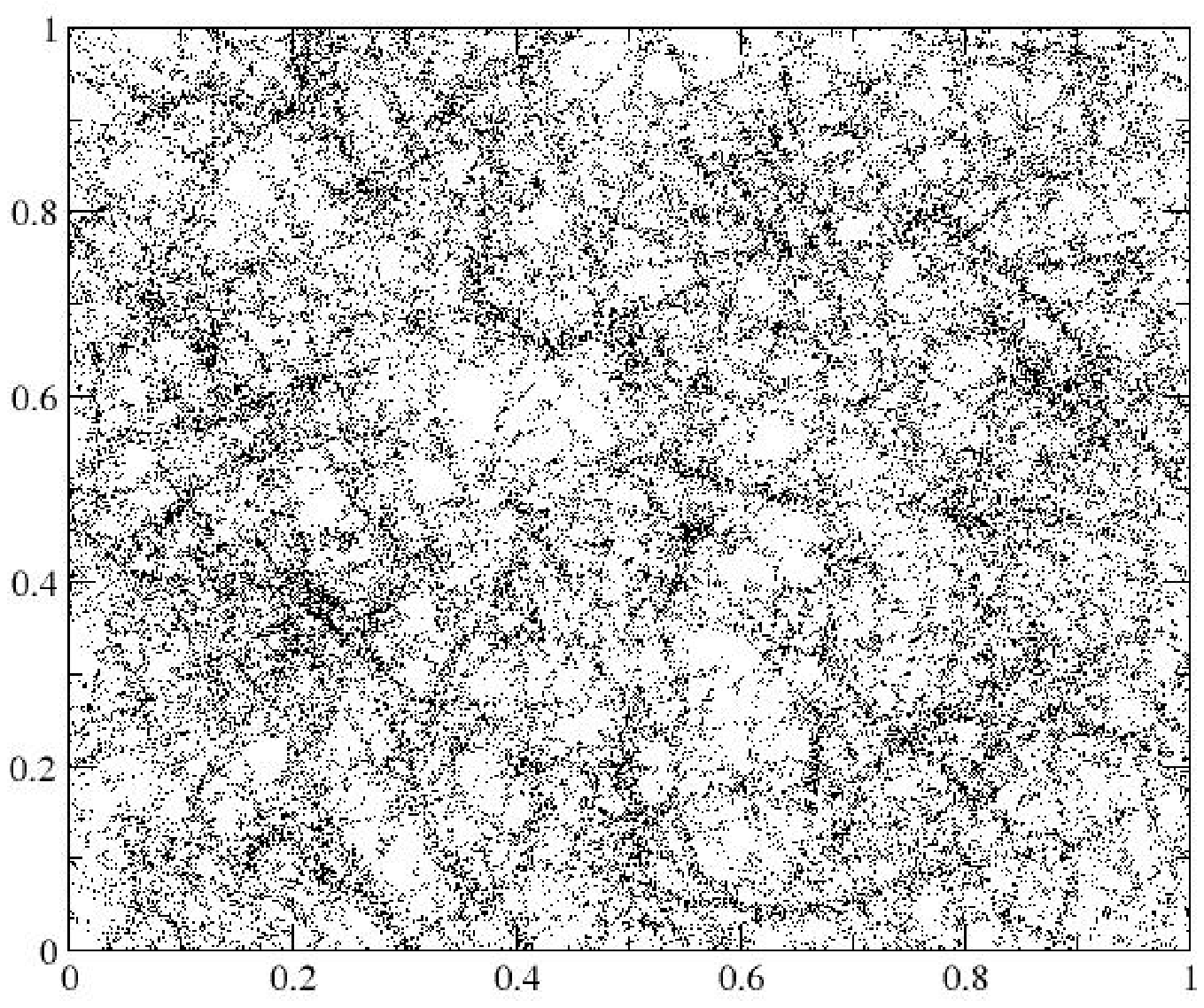}
\includegraphics*[angle=0, width=0.5\textwidth]{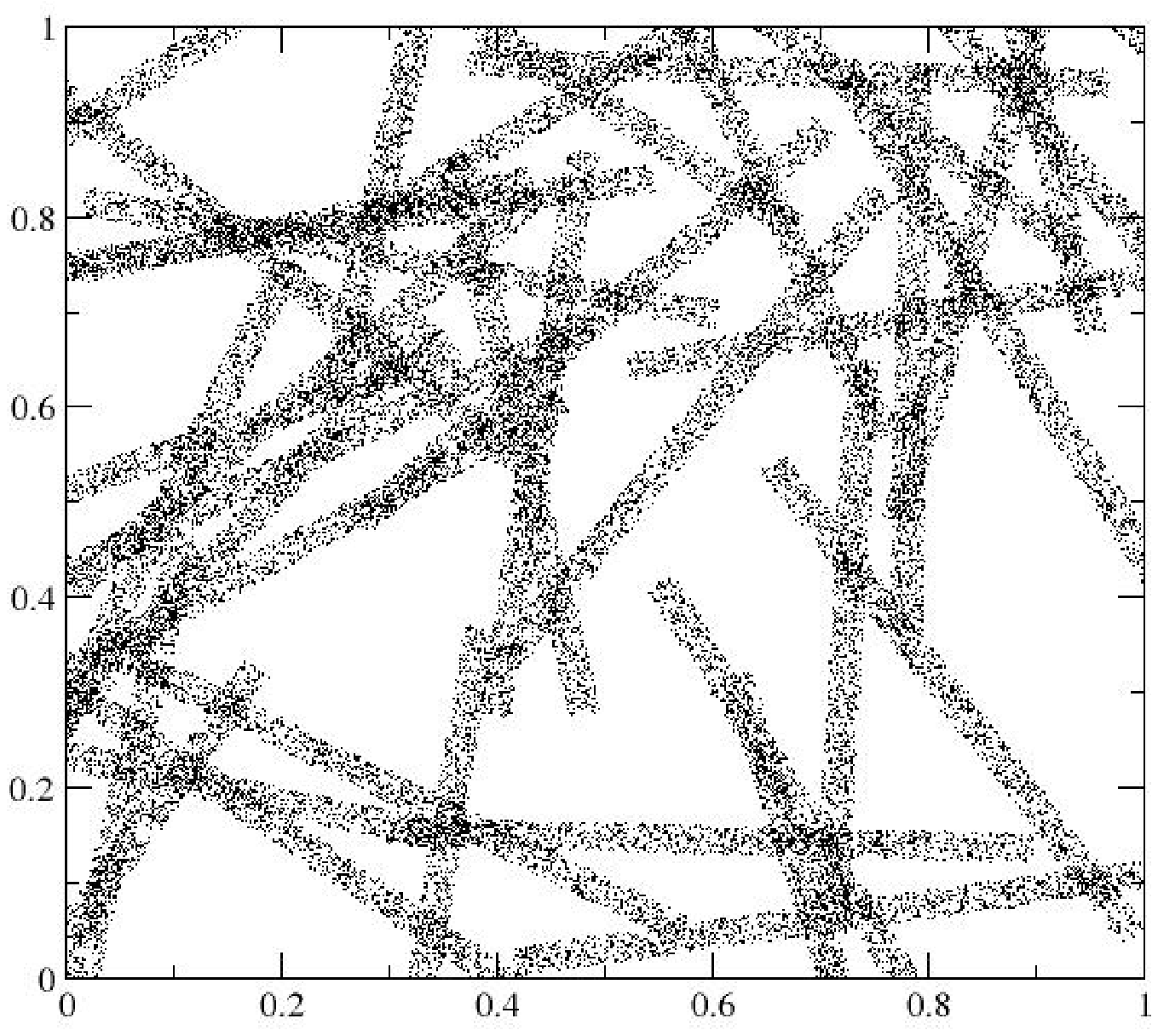}
\includegraphics*[angle=0, width=0.5\textwidth]{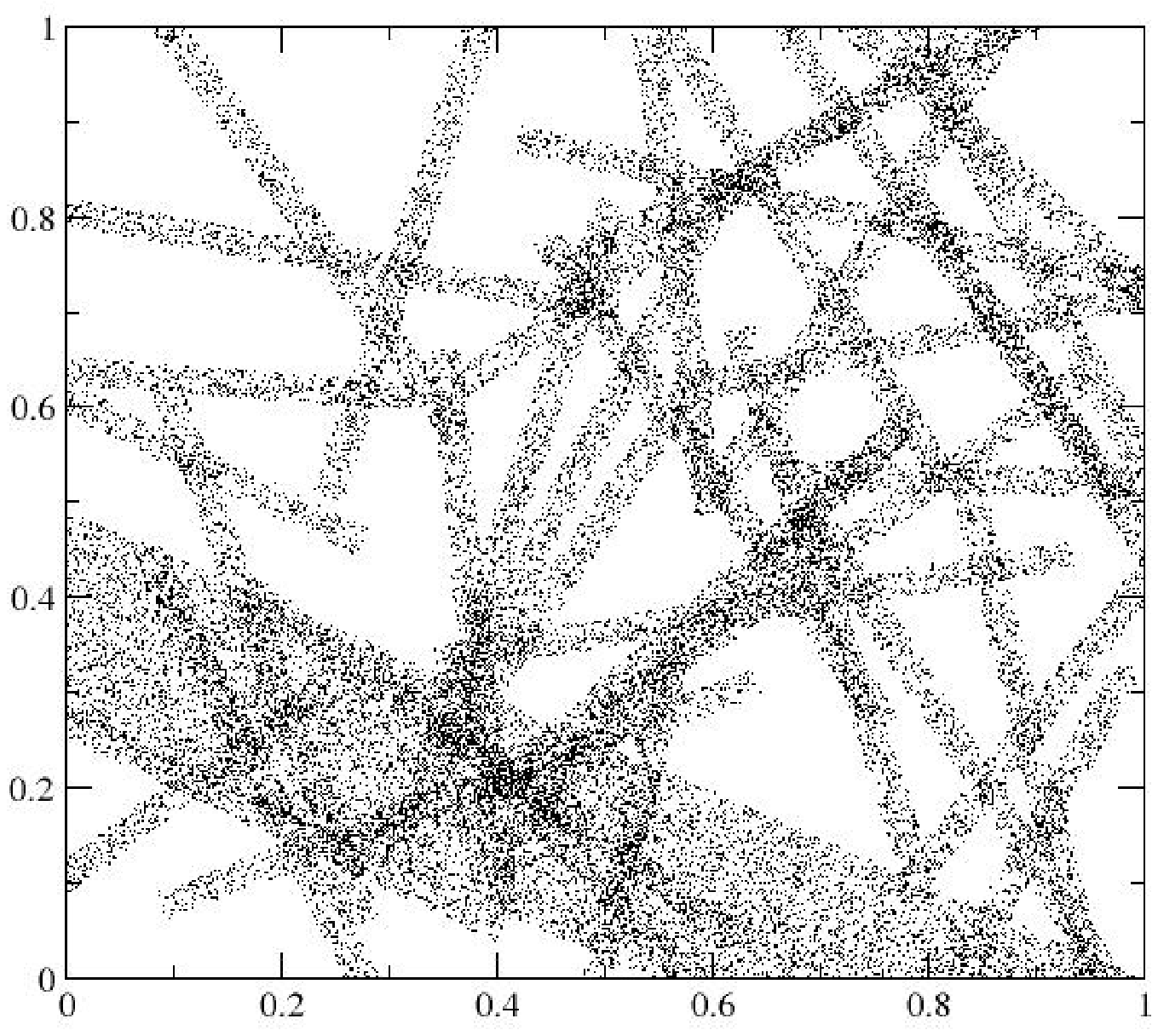}
\caption{Four different realizations of the toy model discussed in the
  text.  Upper-left panel: uniform distribution with short-range
  positive correlations (T0). Upper-right panel: uniform distribution
  with long-range positive correlations (T1).  Bottom-left panel:
  nonuniform distribution with long-range positive correlations (T3)
  Bottom-right panel: nonuniform distribution with long-range
  positive correlations and non self-averaging properties (T4).}
\label{toy} 
\end{figure*}

\begin{figure}
\begin{center}
\includegraphics*[angle=0, width=0.5\textwidth]{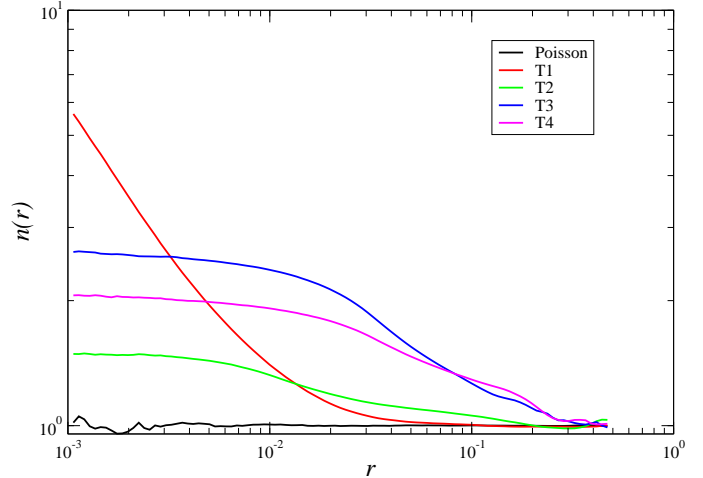}
\end{center}
\caption{Conditional density for the toy models shown in
  Fig.\ref{toy}. The case of a Poisson point distribution is added as
  a reference. (The conditional density has been normalized to the
  number of points in the simulations.) The model T1 has a short-range
  correlation, which corresponds to a fast decay of $\overline{n(r)}$.
  The model T2 is still uniform on large scales, i.e.
  $\overline{n(r)}$ is flat. The models T3 and T4 have strong
  clustering up to the box size.}
\label{fig:gamma_toy}
\end{figure}

\begin{figure}
\begin{center}
\includegraphics*[angle=0, width=0.5\textwidth]{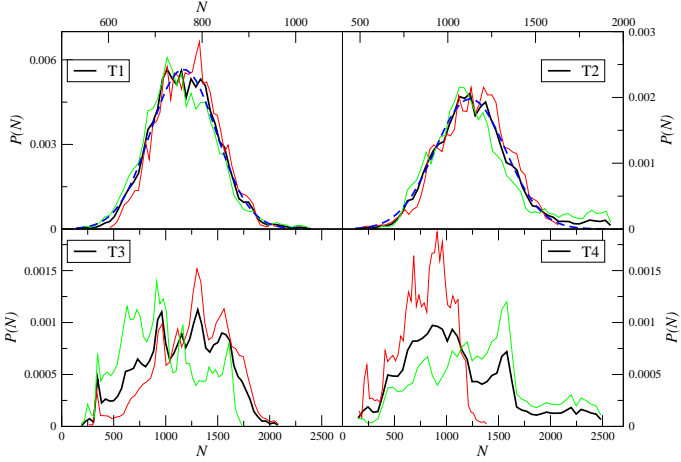}
\end{center}
\caption{PDF of conditional fluctuations (black line) filtered at 1/10
  of the sample size (i.e., $r=0.1$) for the toy models shown in
  Fig.\ref{toy}. Also shown is the PDF computed in two parts of the
  box, i.e. for $y>0.5$ (red line) and $y<0.5$ (green line). Both the
  models T1 and T2 approach to a Gaussian distribution (blue dashed
  lines), as these distributions are uniform although correlated.  The
  PDF of the model T3 does not approach a Gaussian function but it is
  self-averaging inside the box. Finally the PDF of the model T4 is
  not Gaussian and it does not show self-averaging properties.}
\label{fig:pdf_toy}
\end{figure}

}

\subsection{Strategy for a statistical analysis of a finite-sample
distribution}

In a {\it finite sample} we need to set up a strategy for testing
the different assumptions used in the statistical analysis. To this
aim we have  to make a clear distinction between statistical quantities
that are normalized to the sample average density and those that are
not. Given that the primary scope of our study is to determine whether
a statistically meaningful estimate of the average density is possible
in the given samples, we mainly use statistical quantities that do not
require the assumption of homogeneity inside the sample and thus avoid
the normalization of fluctuations to the estimation of the sample
average. These are thus conditional quantities, such as the
conditional density $n_i(r)$ from the $i^{th}$ galaxy, which gives the
density in a sphere of radius $r$ centered on the $i^{th}$
galaxy. Conditional quantities are well-defined both in the case of
homogeneous and inhomogeneous point distributions. If a distribution
is self-averaging inside a given sample or in the range of scales
where such a property is found to hold, then it is possible to
consider the whole sample average of the conditional density, which is
determined by computing
\be
\label{first_moment}
\overline{n(r)} = \frac{1}{M} \sum_{i=1}^{M} n_i(r)
\ee
with respect to the $i=1,...,M$ galaxies contained in the given
sample.  When a distribution is nonuniform (i.e. inhomogeneous), the
conditional variance, which quantifies the amplitude of conditional
fluctuations, is such that
\be
\label{second_moment}
\overline{\delta(r)^2} = \frac{ \overline{n(r)^2} - \overline{n(r)}^2}{
  \overline{n(r)}^2} \sim {\cal O} (1)  \;, 
\ee 
where the last equality corresponds to the fluctuations being
persistent \citep{gsl01}.  On the other hand for homogeneous
distributions, with any kind of small-amplitude correlations we find
that \citep{gsl01}
\be 
\overline{\delta(r)^2} \ll 1 \;.  
\ee

To test whether a distribution is self-averaging inside a given sample
one may measure the PDF of conditional fluctuations and determine
whether this is stable in different subregions of the given
sample. Only when the statistical self-averaging property is satisfied
may one consider determining whole-sample average quantities. Then
only if the conditional density is roughly constant inside a given
sample, and thus the distribution in that sample is approximately
uniform, may one consider determining fluctuations in
amplitude and their correlations normalized to the sample density.

Quantities like the two-point correlation function,  whose
estimator can  generally 
be written as \citep{book}
\[
\overline{\xi(r)} = \frac{\overline{n(r)}}{n_S} -1 \;,
\]
measure the correlation amplitude of fluctuations with respect to 
determining the sample average $n_S$
\footnote{We remind the reader that the previous equation is also
  valid in the infinite volume limit. The various estimators that can
  be found in the literature use different methods to treat boundary
  conditions, hence to estimate both the nominator and the denominator
  of the previous relation \citep{book}.}.  When the distribution is
nonuniform the estimation of the sample average is ill-defined, even
if the distribution is self-averaging inside the sample volume,
resulting in systematic effects in determining the estimator
$\overline{\xi(r)}$ \citep{book}. Thus unconditional quantities are
only well-defined  for uniform distributions.

\section{The samples} 
\label{sec:samples} 

The SDSS \citep{york} is currently the largest spectroscopic survey of
extragalactic objects, and here we consider the data from the public
data release six (SDSS DR6) \citep{paperdr6}
\footnote{{see {\tt www.sdss.org}}} containing redshifts for about
800,000 galaxies and 100,000 quasars.  There are two independent parts
of the galaxy survey in the SDSS: the main galaxy (MG) sample and the
luminous red galaxy sample. We only discuss the former.  The
spectroscopic survey covers an area of 7425 square degrees on the
celestial sphere. The Petrosian apparent magnitude limit with
extinction corrections for the galaxies is 17.77 in the $r$-filter and
photometry for each galaxy is available in five different bands. A
detailed discussion of the spectroscopic target selection in the SDSS
MG sample can be found in \citet{strauss2002}.

\subsection{The query from the SDSS database}

We  used the following criteria to query the SDSS DR6 database, in
particular from the {\tt SpecPhoto} view \citep{strauss2002,paperdr6}:

\begin{itemize}

\item We constrained the flags indicating the type of object so that we select
only the galaxies from the MG sample,  i.e. ({\tt specClass = 2} and 
{\tt (primTarget \& 64) $>$ 0 or   (primTarget \& 128) $>$ 0 or 
 (primTarget \& 256) $>$ 0}).  

\item We constrained the redshift confidence parameter to be $z_{conf}
  \ge 0.35$ with flags indicating no significant redshift
  determination errors, i.e.   {\tt zConf $>$ 0.35 AND zWarning
    \& 193821 =0 AND NOT zStatus IN (0, 1, 2)}.

\item We then considered galaxies in the redshift range $10^{-4} \leq
  z \leq 0.3$, i.e.  {\tt z $>=$ 0.0001 AND z$<=$ 0.3}.  Given the low
  value of the lower redshift limit, nearby galaxies that are large
  enough may get ``shredded'' into smaller pieces by the SDSS
  automatic pipelines and may represent an unwanted contamination to
  the data. However, these are excluded by considering samples that 
  have not a too low redshift limit and that do not contain
  extremely bright galaxies (see below).

\item We applied the filtering condition for Petrosian apparent
  magnitudes with extinction corrections $r < 17.77$, thus taking the
  target magnitude limit for the MG sample in the SDSS DR6 into
  account, i.e.  {\tt (petroMag\_r - extinction\_r) $<$ 17.77}.
\end{itemize}

In this way we  selected 525,813 galaxies.

\subsection{The angular regions} 

We use the internal angular coordinates of the survey $(\lambda,
\eta)$, which can be transformed into the usual equatorial angular
coordinates by a simple rotation. The angular coverage of the survey
is not uniform, but observations have been done in different
disconnected sky regions. For this reason we have considered three
rectangular angular regions in the SDSS internal angular coordinates
in this way we did not have to consider the irregular boundaries of
the survey mask, as we cut such boundaries to avoid uneven edges of
observed regions.  In Table \ref{tbl_VLSamplesProperties2} we report
the limits of the cuts are chosen using the internal coordinates of
the survey $\eta$ and $\lambda$ (in degrees) and the sample solid
angle $\Omega$ in steradians.  We did not use corrections for the
redshift completeness mask or for fiber collision
effects. Completeness varies most near the current survey edges, which
are excluded in our samples. Fiber collisions in general do not
present a problem for measurements of large-scale galaxy correlations
\citep{strauss2002}.
\begin{table}
\begin{center}
\begin{tabular}{|c|c|c|c|c|c|}
\hline
  Region  & $\eta_{min}$ & $\eta_{max}$ 
& $\lambda_{min}$ & $\lambda_{max}$ & $\Omega$ (sr.)\\
\hline
    R1    & -6.0   & 36.0  &-48.0 &  32.5& 0.94 \\
    R2    & -33.5  & -16.5 &-54.0 & -17.0& 0.15 \\
    R3    & -36.0  & -26.5 &-14.0 &  43.0& 0.15 \\
\hline
\end{tabular}
\end{center}
\caption{Properties of the angular regions considered.}
\label{tbl_VLSamplesProperties2}
\end{table}

Let us add a comment on incompleteness, which we concluded does not
play a major role in out results. This conclusion is reached by
considering several fact. 

(i) All statistical quantities we measured, such as counts of galaxies
as a function of apparent magnitude, the redshift distribution in the
magnitude-limed sample, and the measurements of the correlation
function in volume-limited samples, agree very well with previous
works that have taken into account the variation in completeness in
the whole survey area \citep{zehavi_earlydata,zehavietal05}.  This
implies that there are no major differences in the way we treated the
data, while there is a substantial difference in the interpretation of
the results of the statistical analysis, as we discuss below.

(ii) Some authors use the method of making a random catalog with the
same selection function of the real sample, and to this aim the
detailed information given by the survey completeness mask is used.
The completeness mask takes (mainly) into account that the limiting
magnitude has small variations in different fields, so that some
galaxies have not been observed and so that a small fraction of
galaxies in the photometric catalog have not been observed.  There is
not a way, that is free of {\it a-priori} assumptions, to correct for
such an incompleteness.  Given that the detailed information of the
real galaxy distribution is unknown one has to make some assumptions
on the statistical properties of such a distribution.  On the other
hand, a way of checking the possible effects of incompleteness free of
assumptions, is to limit the selection of galaxies to more stringent
limits in apparent magnitude, especially for faint magnitudes. By
limiting the apparent magnitude to 17.5 instead of 17.77, we found no
statistical difference from the results presented in what follows
\citep{paper_sdss1}. A similar conclusion on the survey incompleteness
has been found in the two-degree field galaxy redshift survey (2dFGRS)
\citep{paper_2df_aea}.  In addition it has been shown by \citet{cabre}
that the completeness mask could be the main source of systematic
effects only on small scales, while we are interested in the
correlation function on relatively large separations.

\subsection{The volume-limited samples}

To construct volume limited (VL) samples that are unbiased for the
selection effect related to the cuts in the apparent magnitude, we
applied a standard procedure \citep[see][]{zehavietal05}.
Firstly we computed metric distances as \citep{hoggdistance}
\begin{equation}
\label{MetricDistance} 
R(z; \Omega_m, \Omega_\Lambda) = \frac{c}{H_0}
\int_{\frac{1}{1+z}}^{1} {\frac{dy}{y \cdot
    \left(\Omega_m/y+\Omega_\Lambda \cdot y^2 \right)^{1/2}}} \; ,
\end{equation}
where we used the cosmological parameters $\Omega_m=0.3$ and
$\Omega_\Lambda=0.7$ for the concordance model. We  checked that
our results do not depend significantly on the choice of cosmological
parameters when  taken in a reasonable range of values. This
is expected since the redshift involved in these studies is limited to
$z \le 0.2$ and relativistic redshift-distance corrections are
generally small and linear to the redshift for $z <1$.

Second, the galaxy absolute magnitude was determined to be
\citep[see][]{zehavietal05}
\begin{equation}
\label{AbsoluteMagnitude}
 M_r = m_r - 5  \log_{10}\left[R(z) (1+z)\right] - K_r(z) - 25\;,
\end{equation}
where $K_r(z)$ is the K-correction.

We determined the $K_r(z)$ term from the NYU VACG data 
\footnote{ { see NYU Value-Added Galaxy Catalog, 2008 {\tt
      http://ssds.physics.nyu.edu/} } } \citep{blanton2005}: to
calculate the K-correction, a template fit to observed galaxy fluxes
was used \citep{blanton_roweis2007}.  To match these data with the
results of the query form SDSS data archive we  applied the
following criteria: (i)   right ascension  and declination must match
within 1 arc-sec; (ii) relative difference between redshifts should be
less than $1 \%$.  With these constraints we find 517,729 galaxies
which successfully matched and $8084$ galaxies which are not matched.
For unmatched galaxies we considered a polynomial approximation to
$K_r(z)$
\be 
\label{Kcorr}
K_r(z) = a_0 + a_1 z + a_2 z^2 
\ee
where $a_0 = 0.006$, $a_1=0.847$ and $a_2 = 1.232$. The behavior
of Eq.\ref{Kcorr} corresponds to the average K-correction of matched
galaxies.

As discussed above, the MG sample corresponds to the observations, in
a certain sky area, of all galaxies with apparent magnitude in a given
range. There is thus an intrinsic selection effect because faint
galaxies can only be observed if they are close enough to us, while
brighter galaxies can be observed both at low and high redshift.  Thus
to avoid this observational selection effect, a VL sample is defined
by two cuts in distance and two in absolute magnitude, so that it
covers a rectangular area in the $M-z$ diagram \citep{zehavietal05}.
To define VL samples, we restricted apparent magnitudes to the range
$14.5 \le m_r\le 17.77$, with the bright limit imposed to avoid the
small incompleteness associated with galaxy deblending
\citep{zehavietal05}.  In Table \ref{tbl_VLSamplesProperties1} we
report the limits of the five VL samples we have considered:
$R_{min}$, $R_{max}$ (in Mpc/h) are the chosen limits for the metric
distance; ${M_{min}, \,M_{max}}$ define the interval for the absolute
magnitude in each sample.
\begin{table}
\begin{center}
\begin{tabular}{|c|c|c|c|c|}
  \hline
  VL sample & $R_{min}$ (Mpc/h)& $R_{max}$ (Mpc/h)& $M_{min}$ 
& $M_{max}$ \\
  \hline
    VL1    & 50  & 200 & -18.9 & -21.1   \\
    VL2    & 100 & 300 & -19.9 & -22.0   \\
    VL3    & 125 & 400 & -20.5 & -22.2   \\
    VL4    & 150 & 500 & -21.1 & -22.4   \\
    VL5    & 200 & 600 & -21.6 & -22.8   \\
   \hline
\end{tabular}
\end{center}
\caption{Main properties of the obtained VL samples with K-corrections
  and without E-corrections.}
\label{tbl_VLSamplesProperties1}
\end{table}
In Table \ref{tbl_VLSamplesProperties-VL1} we report the number of
galaxies in each of the three angular regions for the five VL samples:
in the second column there is the case where K-corrections have been
applied, the third column without K+E-corrections and the fourth
column with K+E-corrections.
\begin{table}
\begin{center}
\begin{tabular}{|l|c|c|c|}
  \hline
  VL Sample& Kcorr &None &E+K\\
  \hline
  R1VL1    & 36316 & 35372 & 36693\\
  R2VL1    &  5939 & 5805  & 5992\\
  R3VL1    &  4231 & 4124  & 4290\\
\hline
  R1VL2    & 48745 & 49981 & 53086\\
  R2VL2    & 9805  & 10020 & 10576\\
  R3VL2    & 10363 & 10556 & 11136\\
\hline
  R1VL3    & 58980 & 51039 & 54389\\
  R2VL3    & 11328 & 9738  & 10416\\
  R3VL3    & 11941 & 10410 & 11090\\
\hline
  R1VL4    & 44503 & 33051 & 44276\\
  R2VL4    & 8064  & 5955  & 8044 \\
  R3VL4    & 8057  & 6062  & 8125 \\
\hline
  R1VL5    & 25216 & 20685 & 21707\\
  R2VL5    &  4360 & 3573  & 3786 \\
  R3VL5    &  4113 & 3390  & 3601\\
  \hline
\end{tabular}
\end{center}
\caption{Number of galaxies in each of the VL samples (VL1,...,VL5)
  and in each region (R1, R2, R3). }
\label{tbl_VLSamplesProperties-VL1}
\end{table}

 In what follows we make a detailed study to understand the effects of
 K and of other redshift dependent corrections. The reason these
 corrections could play a role is that they introduce a
 redshift-dependent behavior of secondary quantities (absolute
 magnitude and distance) when they are derived from primary quantities
 (redshift and apparent magnitude). As several statistical quantities
 we discuss in Sect.\ref{sec:sl} show a distance (or redshift)
 dependence, one may ask whether there is an effect from these
 corrections. To constraint the possible effects of these corrections,
 we discuss two different choices of them in
 Sects.\ref{sec:k0}-\ref{sec:Ec}.  We refer to Appendix
 \ref{cosmocorr} for a discussion of the derivation and the role of
 the cosmological corrections.

\subsection{Effect of K-correction}
\label{sec:k0}

To study the effect induced by the K-corrections on the correlation
analysis discussed in what follows, we constructed a set of VL samples
without applying the $K_r(z)$ term in Eq.\ref{AbsoluteMagnitude}.
This choice is clearly not justified from the physical point of view
and can be interpreted as a way to introduce a general linear
redshift-dependent correction to the absolute magnitude-redshift
relation.  The limits in distance of the corresponding VL samples are
the same as for the samples with K-corrections, and the limits in
absolute magnitude are all the same, except for VL2 and VL5 where
there is a difference of 0.1 magnitudes while the range in absolute
magnitudes is the same (see Table~\ref{tbl_VLSamplesProperties1_K0}).
\begin{table}
\begin{center}
\begin{tabular}{|c|c|c|c|c|}
  \hline
  VL sample & $R_{min}$ (Mpc/h)& $R_{max}$ (Mpc/h)& $M_{min}$ 
& $M_{max}$ \\
  \hline
    VL1    & 50  & 200 & -18.9 & -21.1   \\
    VL2    & 100 & 300 & -19.8 & -21.9   \\
    VL3    & 125 & 400 & -20.5 & -22.2   \\
    VL4    & 150 & 500 & -21.1 & -22.4   \\
    VL5    & 200 & 600 & -21.5 & -22.7   \\
   \hline
\end{tabular}
\end{center}
\caption{The same as for Table \ref{tbl_VLSamplesProperties1} but for
  VL samples without K-corrections and without E-corrections.}
\label{tbl_VLSamplesProperties1_K0}
\end{table}
In Table~\ref{tbl_VLSamplesProperties-VL1} we report the number of
galaxies for the five VL samples: one may note that the main changes
occur in the deepest samples (i.e. VL4 and VL5) where the number of
objects decreases by a factor of $\sim 20 \%$.

\subsection{Effect of E-correction}
\label{sec:Ec}

 According to standard models of galaxy formation \citep{kauffmann},
 because of the evolution of stars, elliptical and spiral galaxies
 were more luminous in the past. To take this physical change in the
 galaxy properties into account, one should include some corrections
 to the measured luminosity.  These corrections are generally unknown;
 i.e., there is not an adequate model of evolution to allow for proper
 calculation of the corrected absolute magnitudes.  For this reason
 and because small-scale clustering at low redshift is thought not to
 be affected by galaxy evolution, these corrections have frequently
 been neglected in the construction of VL samples, e.g.,
 \citep[see][]{zehavi_earlydata}.  However, this omission is a
 reasonable working hypothesis {\it only} if one considers local
 (conditional) quantities. Indeed, as we discuss below, when one
 normalizes fluctuations to the sample average, one uses information
 concerning all scales in the sample, so that all statistical
 quantities derived by such a normalization are affected by the
 large-scale properties of the distribution inside the given sample.

As discussed in Appendix \ref{cosmocorr}, the formula $E(z) =
1.6\times z$ has been used more recently as a simple fit for the
average evolution in galaxy luminosities in the recent past
\citep[see][]{tegmark2004,zehavietal05}
\footnote{Note that these authors use $E(z)=1.6(z-0.1)$ to construct
  the absolute magnitude $M_{^{0.1}r}$, corresponding to the SDSS band
  shifted to match its rest-frame shape at $z = 0.1$, from the
  apparent magnitude $m_r$ and redshift $z$ by using
  Eq.\ref{AbsoluteMagnitudeE}. In this case the $K_r(z)$ term is the
  K-correction from the $r$ band of a galaxy at redshift $z$ to the
  $^{0.1}r$ band. Because here we use $K_r(r)$ at $z=0$ instead of at
  $z=0.1$ the evolution correction has to be shifted by 0.1 in
  redshift.}. In this situation the E-corrected absolute magnitude is
\begin{equation}
\label{AbsoluteMagnitudeE}
 M_r = m_r - 5 \log_{10}\left[R(z)(1+z)\right] - K_r(z) - 25 + E(z)\;.
\end{equation}
The limits in distance for the samples with E+K-corrections are the
same as in the K-corrected samples while the limits in absolute
magnitude change (see Table~\ref{tbl_VLSamplesProperties1_EB}). For this
reason a rough comparison of the number of objects in each VL sample
is not meaningful.
\begin{table}
\begin{center}
\begin{tabular}{|c|c|c|c|c|}
  \hline
  VL sample & $R_{min}$ (Mpc/h)& $R_{max}$ (Mpc/h)& $M_{min}$ 
& $M_{max}$ \\
  \hline
    VL1    & 50  & 200 & -18.8 & -21.0   \\
    VL2    & 100 & 300 & -19.7 & -22.0   \\
    VL3    & 125 & 400 & -20.4 & -22.2   \\
    VL4    & 150 & 500 & -20.9 & -22.4   \\
    VL5    & 200 & 600 & -21.4 & -22.6   \\
   \hline
\end{tabular}
\end{center}
\caption{
The same of Tab.\ref{tbl_VLSamplesProperties1} but for 
VL samples with E+K-corrections and without
(see text for details).}
\label{tbl_VLSamplesProperties1_EB}
\end{table}
In Table~\ref{tbl_VLSamplesProperties-VL1} we report the number of
galaxies for the five VL samples.


\section{Scale-length analysis}
\label{sec:sl} 

As discussed in Sect.\ref{sec:methods}, the main stochastic variable
that we consider and for which we determine statistical properties is
the conditional number of points in spheres~\footnote{In Appendix
  \ref{sec:magnlim} we discuss how several properties of galaxy
  fluctuations can be measured in the magnitude-limed sample by
  considering galaxy counts as function of apparent magnitude and the
  redshift distribution. This study has the advantage of using direct
  observational quantities. It is interesting to note that these
  studies are compatible with the results presented in this section.}.
That is to say, we compute for each scale $r$ the $\{ N_i(r)
\}_{i=1...M}$ determinations of the number of points inside a sphere
of radius $r$ whose center is on the $i^{th}$ galaxy.  The number of
centers $M$, as we discuss in more detail below, depends on the sphere
radius $r$, i.e.  $M=M(r)$.  The random variable $N_i(r)$ thus depends
on scale $r$ and on the spatial position of the sphere's center. We
can express the $i^{th}$ sphere center coordinates with its radial
distance $R_i$ and with its angular coordinates $\vec{\alpha}_i =
(\eta_i,\lambda_i)$. Thus, in general, we can write
\be 
\label{eq1a}
N_i(r) = N(r; R_i, \vec{\alpha}_i) \;. 
\ee
 When we integrate over the angular coordinates $\vec{\alpha}_i$ for
 fixed radial distance $R_i$, we find that $N_i(r)=N(r; R_i)$; i.e.,
 it depends on two variables the length scale of the sphere $r$ and
 the distance scale of the $i^{th}$ sphere center $R_i$, so it has
 been called the scale-length analysis \citep{paper_sdss1}.


\subsection{Number of centers as a function of scale} 
\label{sec:centers}
The reason the number of centers $M(r)$ depend on the scale $r$
follows.  The sample geometry is a spherical portion delimited by the
minimal and maximal value of the radial distance and by the angular
coordinates reported in Table~\ref{tbl_VLSamplesProperties2}.  For the
$i^{th}$ galaxy, with coordinates ($R_i,\vec{\alpha}_i)$, we compute
the six distances from the boundaries of the sample and  consider
the minimal one $r_{bc}^i$.  By simple geometrical considerations
these distances are
\bea && 
r_1^i= R_i \cos(\lambda_i) \sin(\eta_i-\eta_{min})
\\ \nonumber 
&& r_2^i=R_i \cos(\lambda_i) \sin(\eta_{max}-\eta_i) 
\\ \nonumber
&& 
r_3^i=R_i \sin(\lambda_i-\lambda_{min}) 
\\ \nonumber
&&
r_4^i=R_i \sin(\lambda_{max}-\lambda_i) 
\\ \nonumber 
&& r_5^i=R_i-R_{min} 
\\ \nonumber
&& 
r_6^i=R_{max}-R_i 
\eea
and $r_{bc}^i = \mbox{min}[r_1^i,r_2^i,r_3^i,r_4^i,r_5^i,r_6^i]$.  The
length scale $r_{bc}^i$ corresponds to the radius of the largest
sphere, which is centered on the position of the $i^{th}$ galaxy and
which is fully contained in the sample volume. As in Eq.\ref{eq1a} we
consider only fully enclosed spheres in the sample volume, then the 
$i^{th}$ galaxy will not be included in $M(r)$ as long as the sphere
radius is $r>r_{bc}^i$. In this situation for large sphere radii,
$M(r)$ decreases and the location of the galaxies contributing to
$M(r)$ is mostly placed at radial distances in the range $\sim
[R_{min}+r$, $R_{max}-r]$ from the radial boundaries of the sample at
$[R_{min}$, $R_{max}]$.

One could also make the choice considering incomplete spheres,
i.e. spheres that are only partially contained in the sample volume.
In this case, one could then weight the number of points inside the
incomplete sphere by the volume of it contained in the sample, thus
obtaining a more robust statistics, especially on large scales. We
avoid this for the following reason. Suppose that outside the sample
there is a large-scale structure (or a deep under-density): the
weighting above will underestimate (or overestimate) the real number
of points inside the full sphere with respect to the incomplete
one. This inevitably introduces a bias in the measurements, which
affect large-scale determinations.  As it is precisely the scope of
our study to determine { the properties of} large spatial
fluctuations, we avoid using a method that implicitly assumes that
these are irrelevant \citep{book,paper_2df_aea}.

Given their different sizes, the number of centers as a function of
scale $M(r)$ is quantitatively different in each of the five VL
samples. However, one may note from Fig.\ref{Centers_ALL} that the
behavior of $M(r)$ is similar in the different cases. For small sphere
radii almost all galaxies are included; i.e., $M(r)$ is equal to the
number of points contained in the sample. Instead, when the sphere
radius becomes comparable to the size of the largest sphere radius
which is fully contained in the sample volume, $M(r)$ shows a fast
decay. The scale at which this occurs, grows proportionally (taking
the sample solid angle fixed) to the depth of the VL sample.  The
largest scales explored in this survey, i.e. $r\approx 100$ Mpc/h, can
only be reached with the deepest VL samples.

\begin{figure}
\begin{center}
\includegraphics*[angle=0, width=0.5\textwidth]{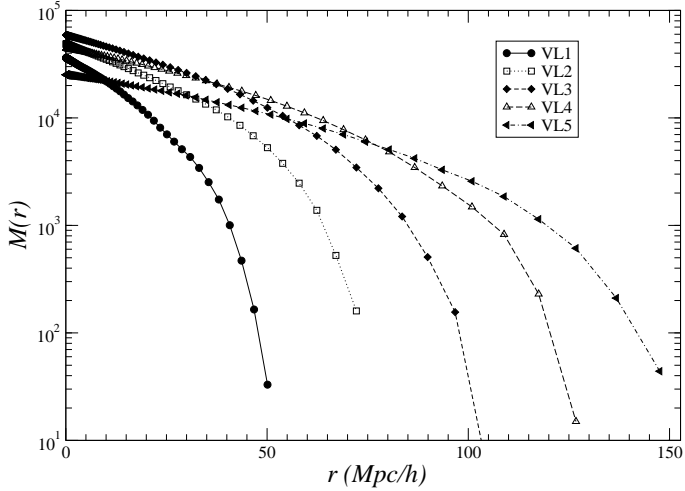}
\end{center}
\caption{Number of centers $M(r)$ as a function of scale in the five
  VL samples (see text for details).}
\label{Centers_ALL}
\end{figure}

\subsection{Probability distribution of conditional fluctuations}

The main information about the statistical properties of the random
variable $N_i(r)$ is provided by its PDF, $P(N,r)$. This gives the
probability distribution to find $N$ points in a spherical volume of
radius $r$ {\it centered on a distribution point}.  It should be
noticed that this is different from the PDF of unconditional
fluctuations, which provides the probability density that in a
spherical volume of radius $r$ {\it centered on an arbitrary point of
  space}, there are $N$ points \citep{saslaw}.  Only when
unconditional properties are well-defined then does PDF of conditional
and the unconditional give similar results \citep{book}.

The frequency distribution in bins of conditional fluctuations at
fixed scale $r$ gives an estimation of the PDF at that scale. The
error bars are computed as the square root of the number of points in
each bin. To compare the behavior in different VL samples, which are
defined by different luminosity cuts so generally containing galaxies
of different absolute magnitudes, we define the normalized variable
\be
\label{pdf_norm1}
x_i(r) = \frac{N_i(r) - \overline{N(r)}}{\overline{\Sigma(r)}} \;,
\ee
and we determine its PDF, that is. 
\be
\label{pdf_norm2}
p(x,r)=P\left(N(r)=\overline{N(r)}+x \overline{\Sigma(r)}\right)
\times \overline{\Sigma(r)} \;,
\ee
where $P(N(r))=P(N,r)$ is the PDF of the variable $N_i(r)$,
$\overline{N(r)}$ is its estimated whole sample first moment and
$\overline{\Sigma(r)}$ is the estimated standard deviation on the
scale $r$.

In Fig.\ref{PDF_VL1_E2K_K0} we show the PDF, estimated in the region
R1 only, of the VL samples with K-corrections, of the samples where
E+K corrections have been applied and finally of the samples in which
no corrections have been imposed.  In Fig.\ref{PDF_VL1_E2K_K0b} we
also show, but only for some cases, the PDF with the estimated Poisson
error bars, together with the best fit obtained by a Gaussian function.
{\it The PDF is not affected by E and/or K
  corrections even in the deepest samples as VL4 and VL5.}  For this
reason, and given that E-corrections are not well-defined, as
discussed above, in what follows we  mostly focus on the case
where only K-corrections have been applied.

\begin{figure*}
\includegraphics*[width=18cm,height=18cm ]{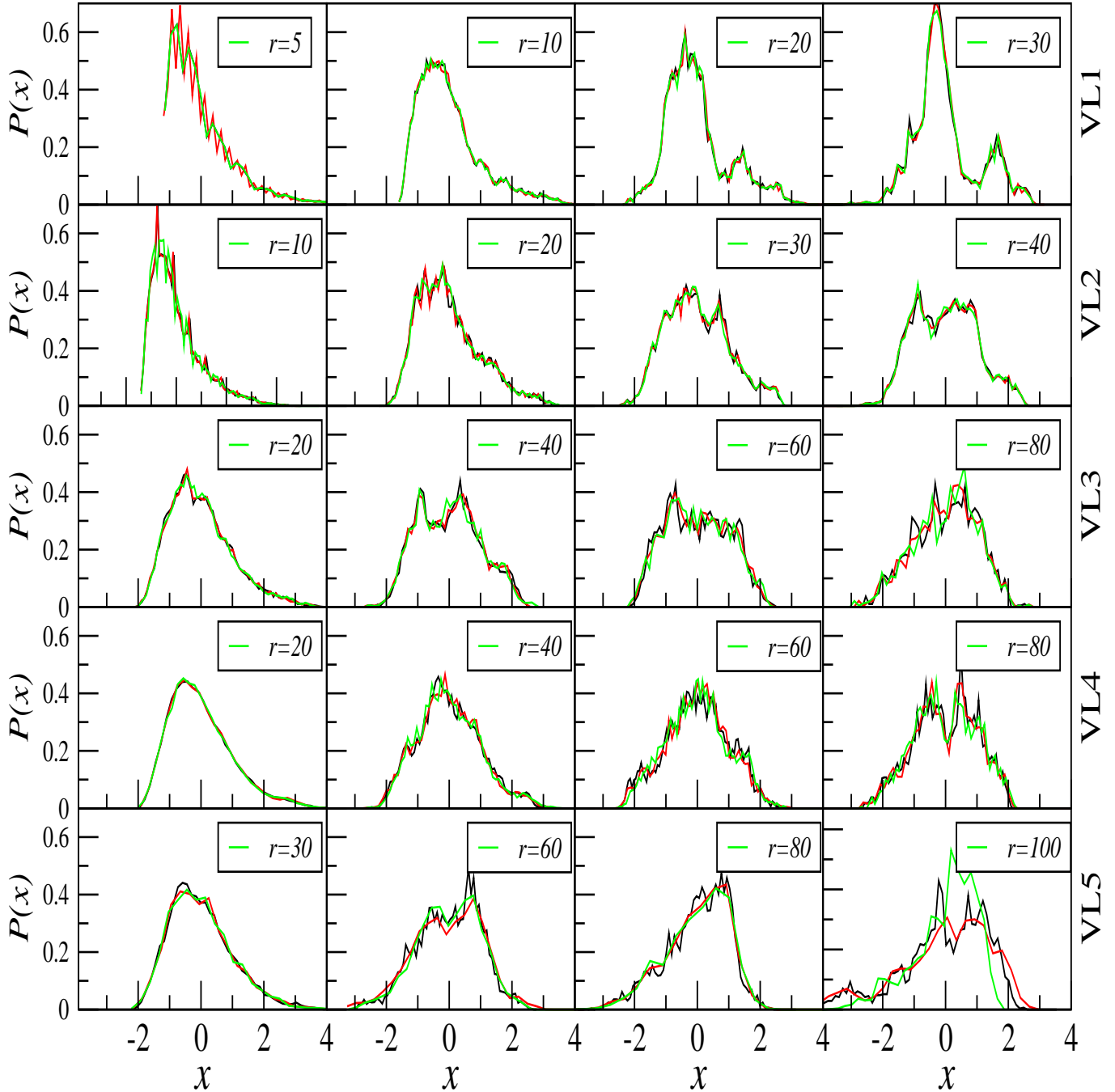} %
\caption{Conditional PDF on different scales for the 5 VL samples
  (each row corresponds to a VL sample; the scale $r$ is reported in the
  caption) with K corrections (black), with K+E corrections (red) and
  without K+E corrections (green).}
\label{PDF_VL1_E2K_K0}
\end{figure*}

\begin{figure}
\includegraphics*[angle=0, width=0.5\textwidth]{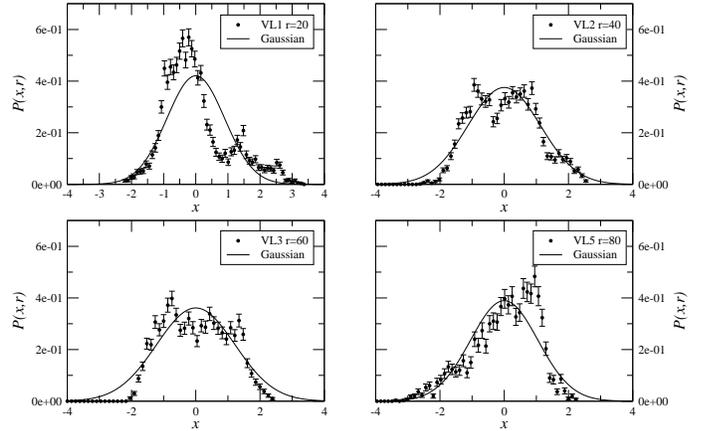} %
\caption{The PDF in
  different samples (with K-corrections only) and for different sphere
  radius with the best-fit Gaussian function (see captions). Poisson
  error bars are reported as a reference.}
\label{PDF_VL1_E2K_K0b}
\end{figure}

It is interesting to compare results for $r=5, 10, 20, 30$ Mpc/h in
different K-corrected VL samples (see Fig.\ref{PDF_ALL_05}): the PDFs
collapse fairly into one another \footnote{The PDF of VL1 for $r=30$
  Mpc/h is not as regular as the other cases because of poor
  statistics.}.
{\it The overall shape is characterized by a long (or fat) tail,
  slowly decaying, for $x$ values high, which makes it substantially
  different from a Gaussian function.}
This is the effect of the large structures (i.e. large fluctuations)
contained in these samples. Similar behaviors have been found in the
2dFGRS \citep{paper_2df_prl,paper_2df_aea}.
\begin{figure*}
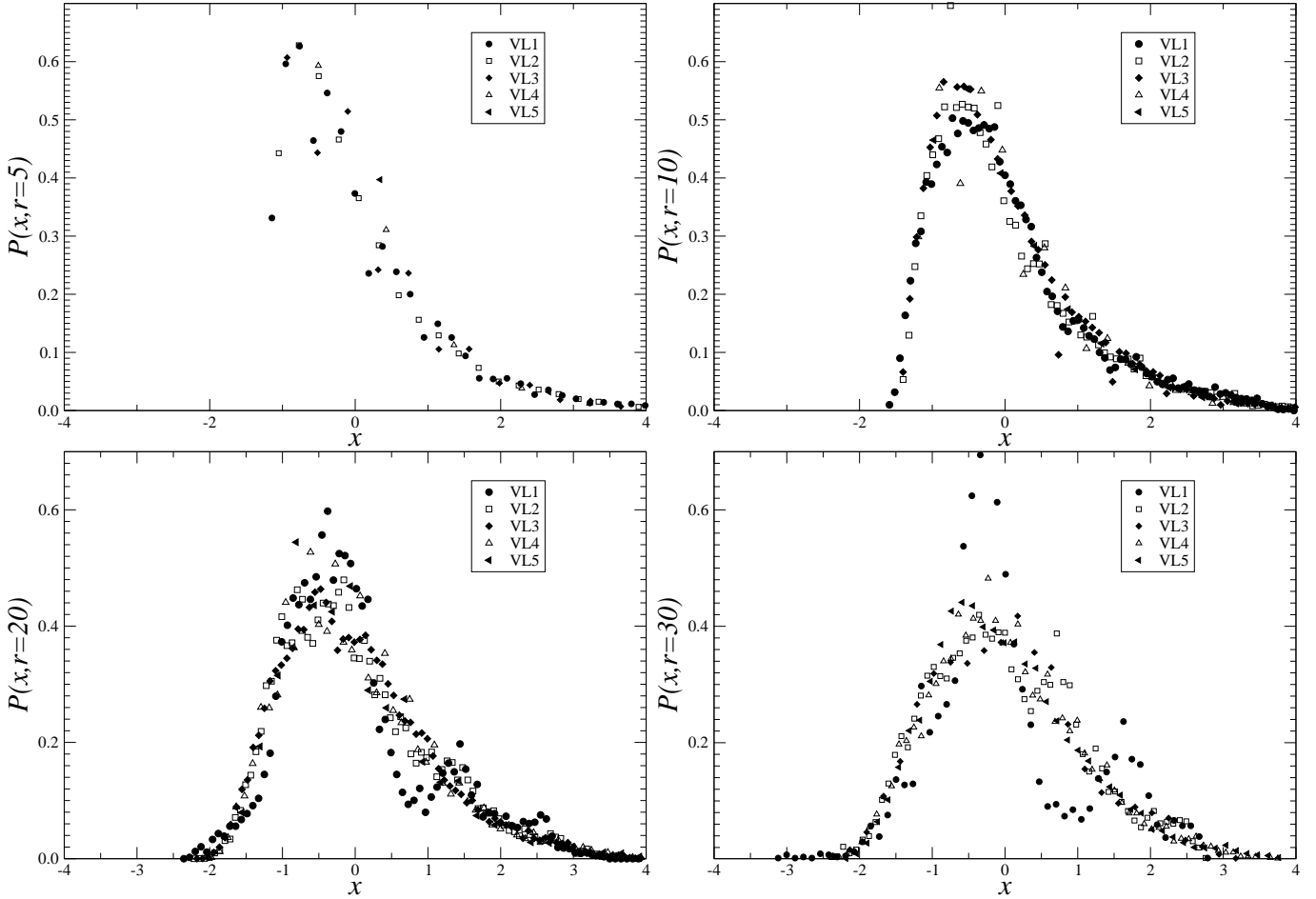

\includegraphics*[angle=0, width=0.5\textwidth]{Fig7a.eps}
\includegraphics*[angle=0, width=0.5\textwidth]{Fig7b.eps}
\includegraphics*[angle=0, width=0.5\textwidth]{Fig7c.eps}
\includegraphics*[angle=0, width=0.5\textwidth]{Fig7d.eps}
\caption{Normalized PDF (see Eq.\ref{pdf_norm1}-\ref{pdf_norm2})
   for $r=5,10,20,30$ Mpc/h in the five VL K-corrected
  samples.}
\label{PDF_ALL_05}
\end{figure*}

Both for small (i.e., $r<30$ Mpc/h) and large (i.e., $r>30$ Mpc/h) the
PDF does not even converge to a Gaussian function.  Actually, for the
largest sphere radii (i.e. $r=80, 100$ Mpc/h), in the sample VL4 and
VL5, the PDF shows a relatively long tail for low $x$ values followed
by a sharp cut-off at values higher than the peak of the PDF.  We
interpret this behavior as due to an intrinsic bias, because given the
finiteness of the sample volume, only a few structures can be
contained in it and thus this statistical measurement cannot properly
give a reliable estimate of large-scale fluctuations.  Already in the
less distant samples (e.g., VL1, VL2, and VL3), the main trends discussed
above are clearly present up to sphere radii $r \approx 80$ Mpc/h.
The distant samples (e.g., VL4 and VL5),  where the effect of other
cosmological corrections maybe more important, allow us to reach the
scales of $\sim 100$ Mpc/h.

To summarize the main result: (i) the PDF is not affected by E and/or
K corrections. (ii) For scales on which conditional fluctuations are
self-averaging and the PDF is stable in different sample subregions,
i.e. for $r<30$ Mpc/h, the overall shape of the PDF is characterized
by a long (or fat) tail that makes it substantially different from a
Gaussian function. (iii) For $r>30$ Mpc/h, the PDF does not converge
to a Gaussian function and it has a different shape in different
samples.  In the next section we present specific measurements to
study the large-scale properties of conditional fluctuations in these
samples testing self-averaging properties.


\subsection{Test for statistical self-averaging}

To study the origin of the differences in the behavior of the PDF in
different VL samples, for large enough sphere radii, we can consider a
specific test. This is useful for studying the self-averaging
properties of the distribution in a given sample.  This test allows us
to establish whether, inside a given sample, it is meaningful to
derive, for instance, whole-sample average quantities and whether we
can consider that a certain estimator gives a reliable and stable
measurement of the ensemble properties of the distribution.

We divide the sample volume into two nonoverlapping subvolumes of
same size, one near of volume $V_n$, and the other more distant of
volume $V_f$, and we determine whether statistical quantities are
stable or show systematic differences in these subsamples. In
principle the ideal test would be to compute the PDF in many different
and nonoverlapping subvolumes, more than the two we use here. The
limitation we face in doing this stems from only the data available in
the SDSS-DR6 and the corresponding sample volumes. In the future data
releases, once the regions R1, R2, and R3 will become contiguous, we
will be able to consider more subvolumes of a single sample.

Given the two limits of the sample in radial distance, $R_{min}$ and
$R_{max}$, we computed the distance $R_h$ at which $V_n = V_f$, thus
obtaining
\be 
R_h = \left(\frac{R_{max}^3-R_{min}^3}{2}\right)^{1/3} \;.
\ee 
To increase the statistics, for a large enough sphere radius $r$, we
have allowed the center of a sphere of radius $r$ to be at a distance
$d$ from $R_h$ such that $r>d$. In this situation the sphere, whose
center is placed, e.g., in the less distant subsample, has part of its
volume in the more distant subsample and vice-versa. Thus a certain
overlap of the determinations of $N_i(r)$ is allowed between the two
half-regions.  This method gives a conservative estimate of the actual
fluctuations between the subsamples. Indeed the overlapping of
different determinations clearly smooths out fluctuations between the
two subsamples: thus any difference we find is certainly a genuine
feature of the distribution.

In addition for each VL sample we consider the PDF determined by all
the values, at fixed $r$, in all three sky regions. The determination
of $N_i(r)$ has to be done separately, for each VL sample, in the
three different sky regions R1, R2, and R3 because of the geometrical
constraints discussed above. This allows us to improve the statistics,
although the R1 region contains about a factor 10 more galaxies than
the other two regions and its larger volume allows many more
determinations than in the two other regions. In particular for a
large enough sphere radius only the values in the R1 region can be
measured.

Results for K-corrected samples are shown in Fig.\ref{PDF_VL1}. The
peak of the PDF in the two half volumes of the different VL samples is
located approximately at the same $N$ value for $r \le 30$ Mpc/h:
although in this range of sphere radii a difference is sometimes
detectable in the location of the peak (e.g., in the samples VL4 and
VL5), the overall shape of the PDF does not substantially change in
the two subvolumes; instead for $r>30$ Mpc/h, the whole PDF shows a
systematic shift, because the shape is very sensitive to the different
kinds of fluctuations (structures) present in each subvolume.  In this
situation the estimation of the first and second moment in the whole
sample is affected by systematic effects that preclude a statistically
meaningful information from them.

In all samples but VL2, the PDF is shifted more toward lower $N$
values in the nearby part of the sample than in the more distant one.
This occurs because fluctuations are generally wilder in the more
distant part of the sample.  This is the effect of the sample
geometry: larger structures can only be found  where the geometry of
the sample volume allows to them contain  and indeed this happens
toward the far  boundaries of the samples.

The sample VL2, for $r> 20$ Mpc/h, shows an interesting and peculiar
feature: particularly, the PDF in the nearby subvolume is shifted
toward higher $N$ values than that in the more distant one.  In this case,
there is a large under-dense region for $R>220$ Mpc/h extending up to
the limits of the sample at $R=300$ Mpc/h (see discussion below).  The
trend found in VL2 is interesting, because it shows that there is not
only the occurrence of large fluctuations in the more distant part of
the sample volume, which could be thought to be ascribed to a
systematic selection effect other than structures. {\it It shows
  instead that there is not such a systematic trend in each of the
  samples.}

This situation  clearly agrees  with the behavior of the whole
sample PDF discussed in the previous section, particularly that there
are, at the same sphere radius $r$,  detectable changes in shape of the
PDF in different VL samples.  This implies that the sample volumes are
not large enough to allow  stable determination of the PDF and its
moments for sphere radii $r>30$ Mpc/h.

As a final remark, to reach the important conclusion about non
self-averaging properties of conditional fluctuations, when they are
filtered on scale $r>30$ Mpc/h, it is enough to consider the nearby
samples VL1, VL2, and VL3. In these samples, due to the narrow range
of redshifts involved, any other type of cosmological correction than
the ones considered here, is expected to perturb our results a little.

\begin{figure*}
\includegraphics*[width=18cm,height=18cm]{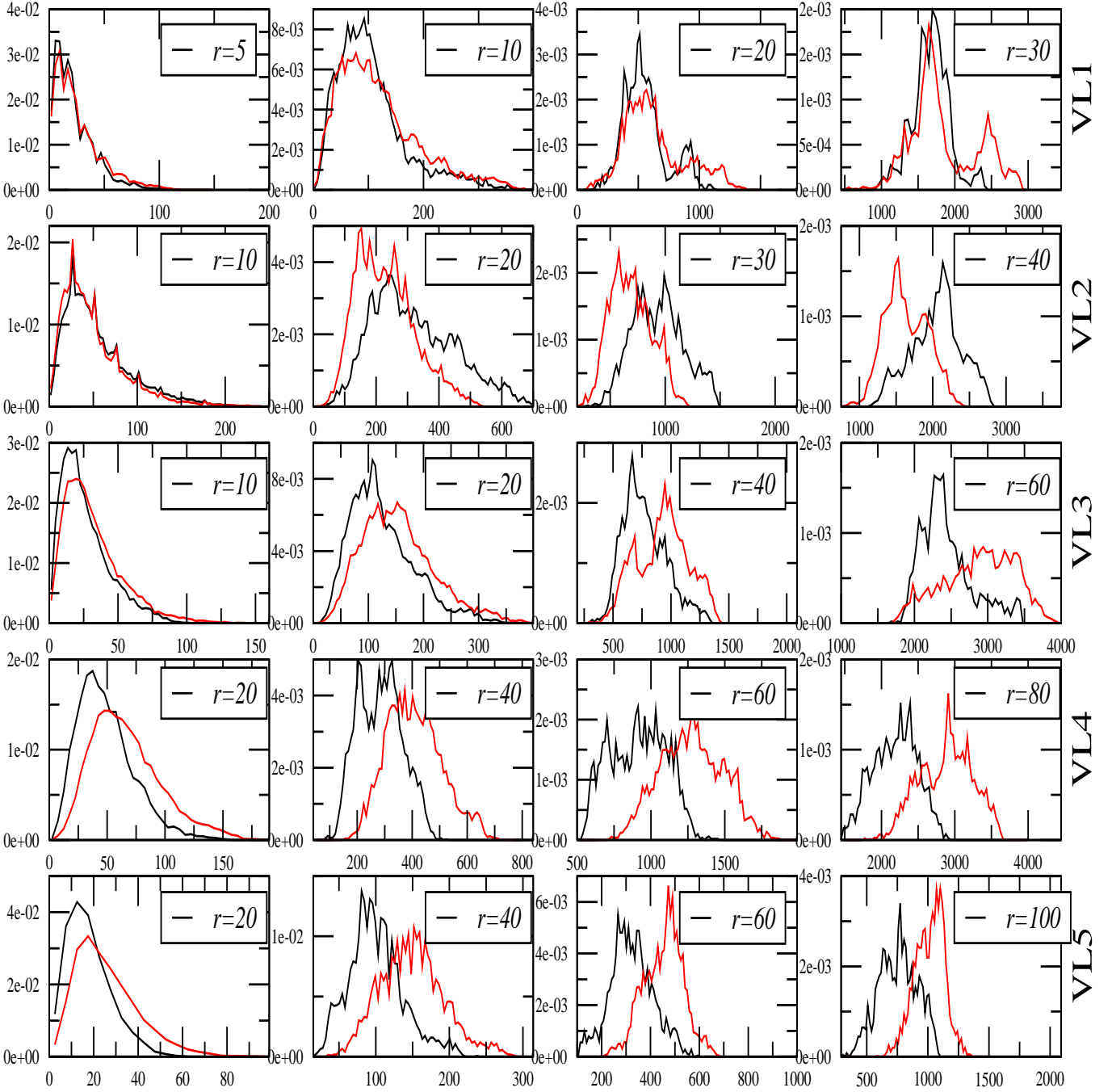}
\caption{PDF in the two subvolumes of the K-corrected VL samples (each
  row corresponds to a VL sample): the black line marks the PDF in
  nearby subsample, and the red line in the more distant subsample. The
  $x-$axis reports the number of points $N(r)$ (the scale $r$ is
  reported in the caption) and the PDF $P(N;r)$ is on the $y-axis$.}
\label{PDF_VL1}
\end{figure*}

\subsection{Effect of K-corrections and evolutionary corrections} 

As illustrative examples of the situation in the samples with E+K
corrections, and in those where no corrections are applied at all, we
show in Fig.\ref{PDF_VL5a_EK} the cases of VL3 and VL5. In the former
one the corrections, because of the relatively high redshifts
involved, are expected to modify the behaviors more.  As one can see
from the above figures, there is no substantial change with respect
to the case where only K-corrections are applied.  Thus even in this
case, the effect of K+E corrections represents minor modifications to
the measured behaviors.

\begin{figure*}
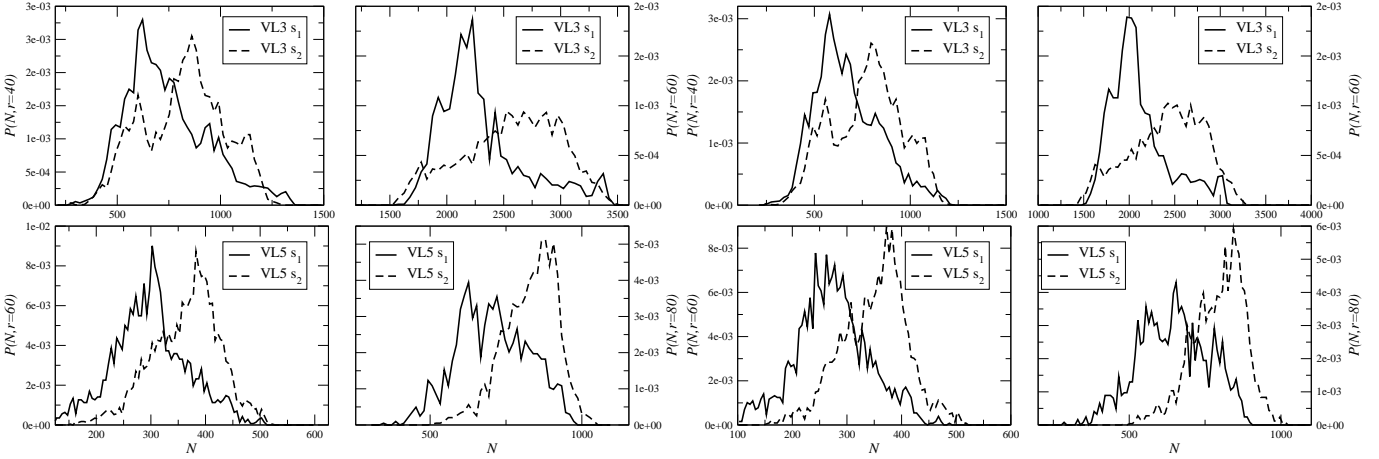

\includegraphics*[angle=0, width=0.5\textwidth]{Fig9a.eps}
\includegraphics*[angle=0, width=0.5\textwidth]{Fig9b.eps}
\caption{As Fig.\ref{PDF_VL1} but for the K+E-corrected VL3 and VL5
  samples (left) and for the same samples  without
  K+E corrections (right).  }
\label{PDF_VL5a_EK}
\end{figure*}


\subsection{Average in bins} 
\label{sec:ave_bin}

\begin{figure*}
\includegraphics*[angle=0, width=0.5\textwidth]{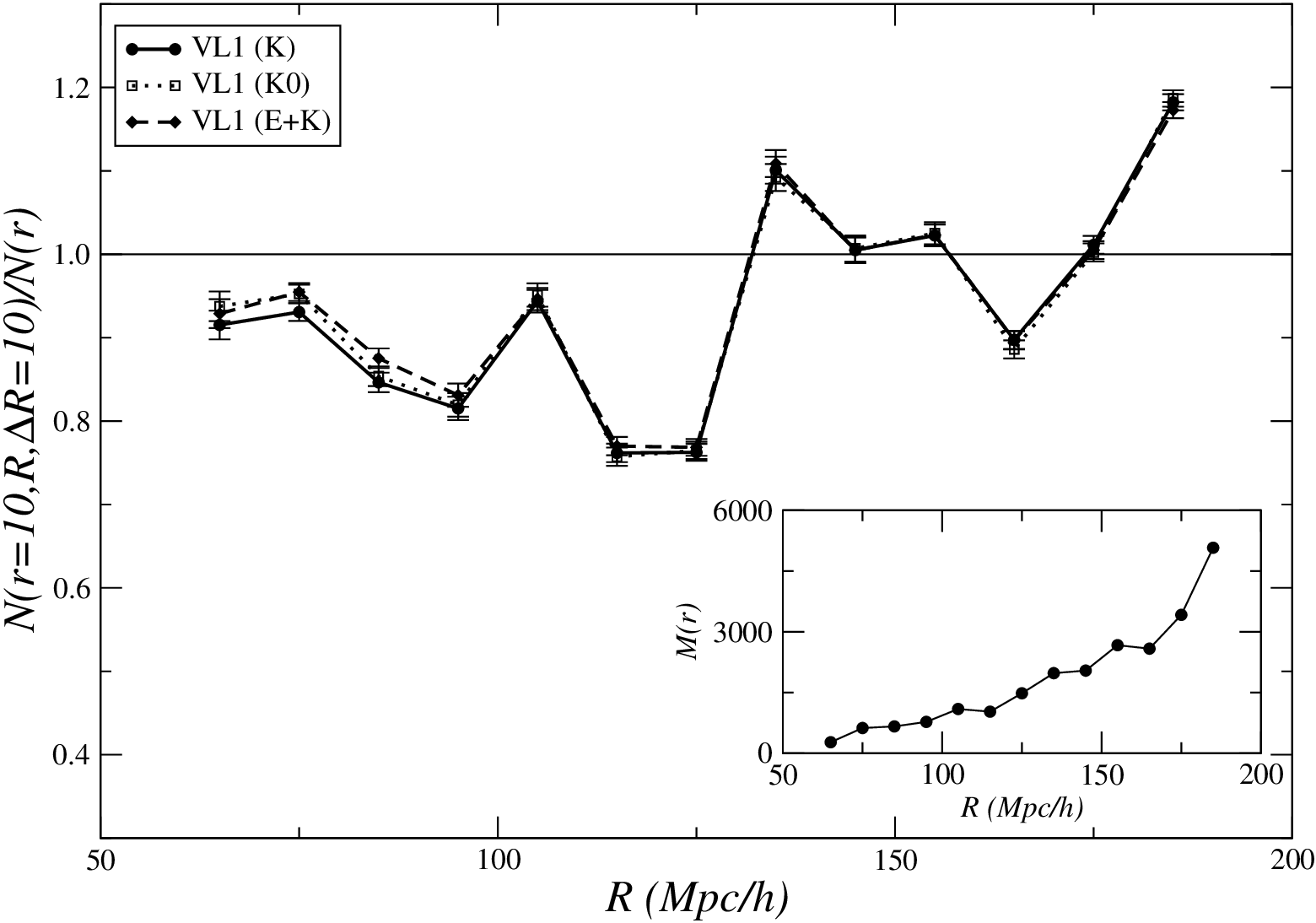}
\includegraphics*[angle=0, width=0.5\textwidth]{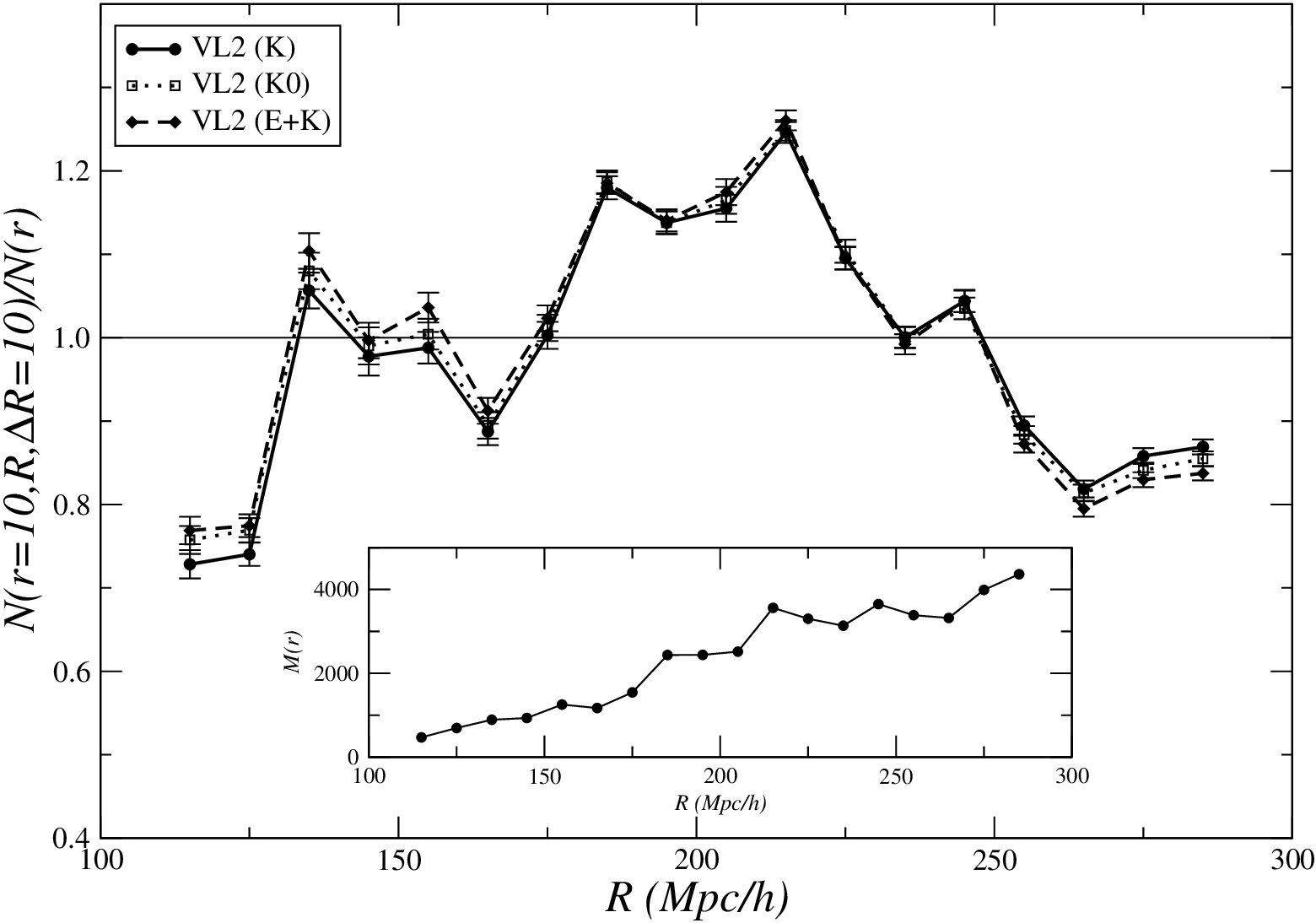}
\includegraphics*[angle=0, width=0.5\textwidth]{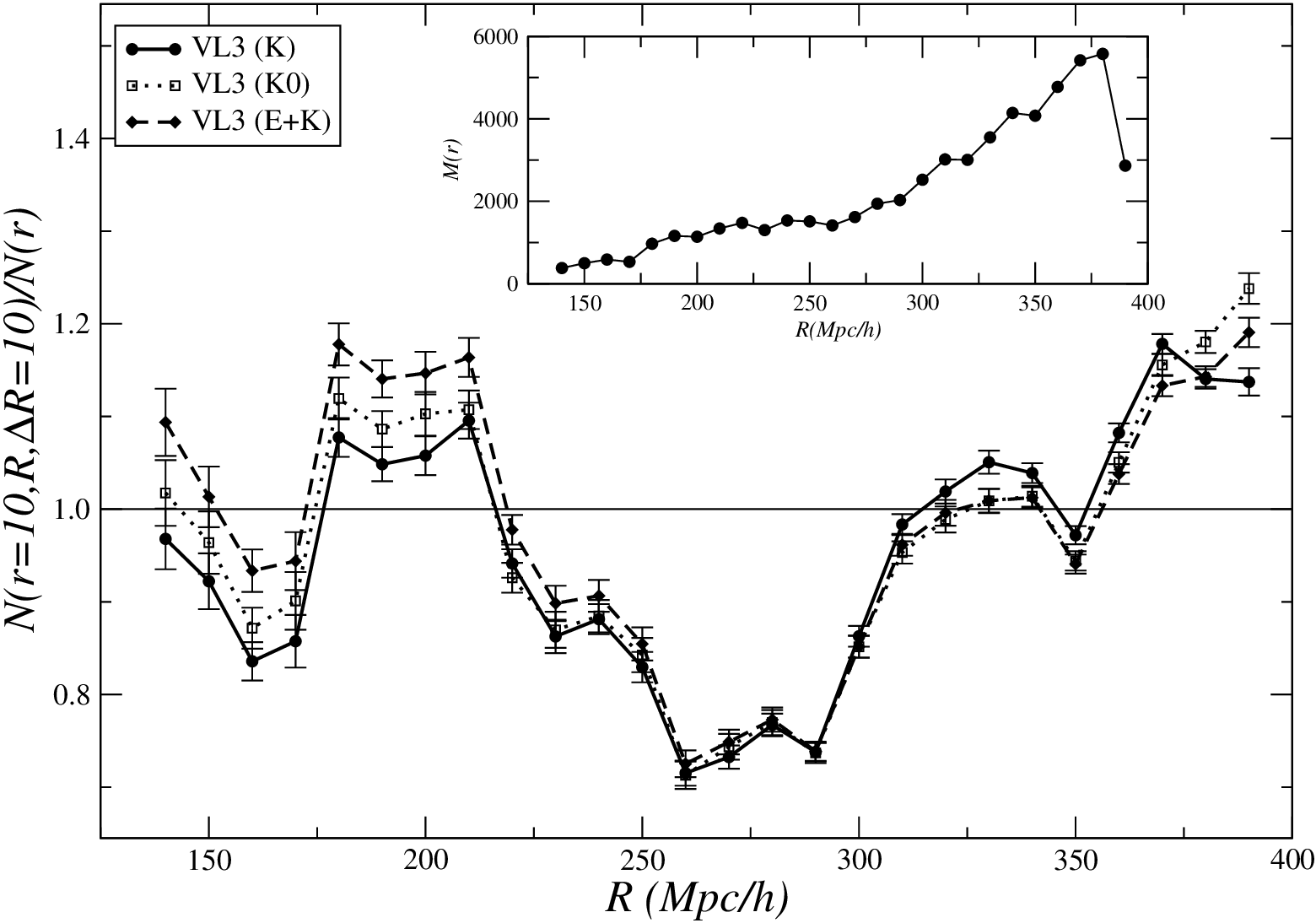}
\includegraphics*[angle=0, width=0.5\textwidth]{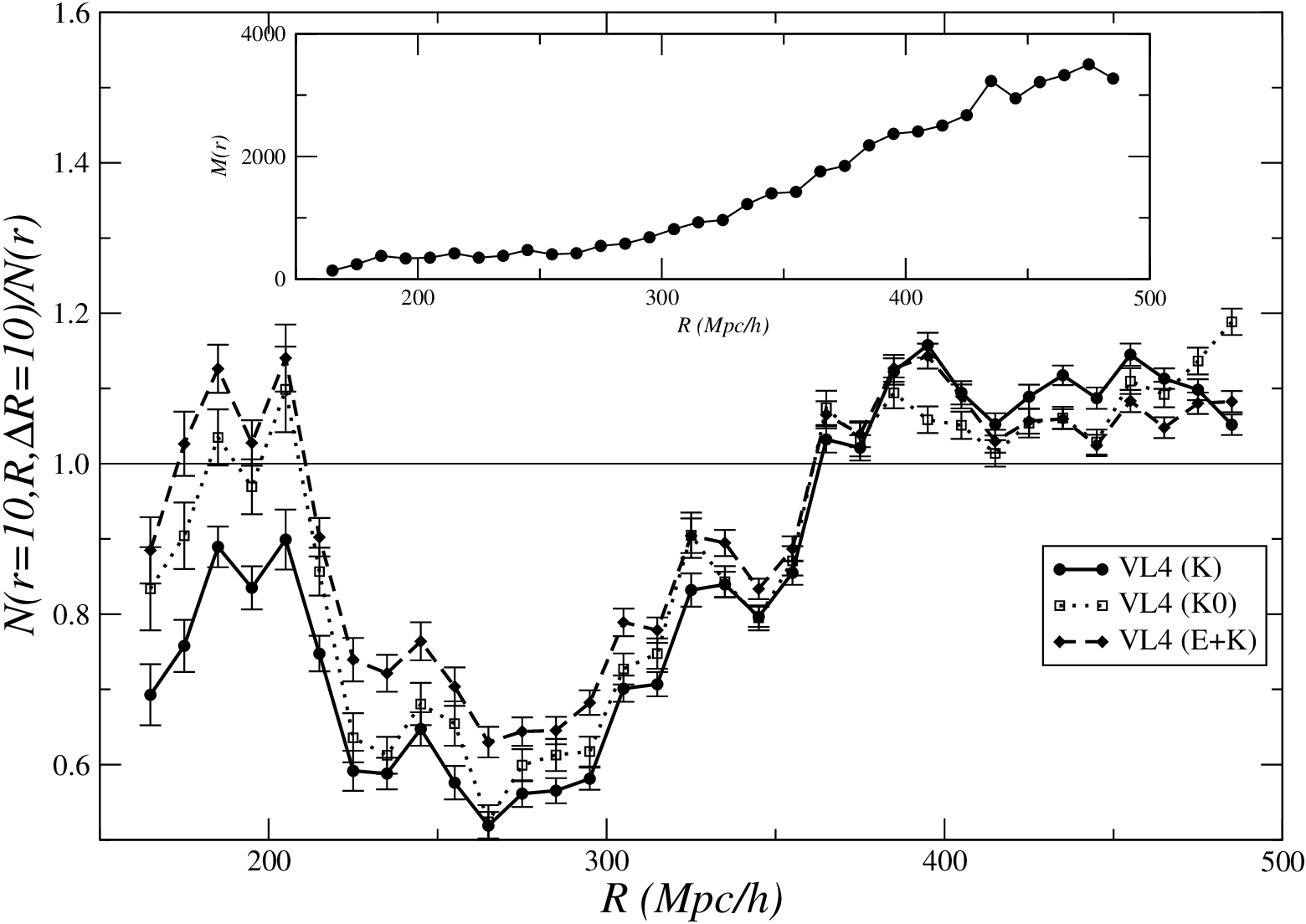}
\includegraphics*[angle=0, width=0.5\textwidth]{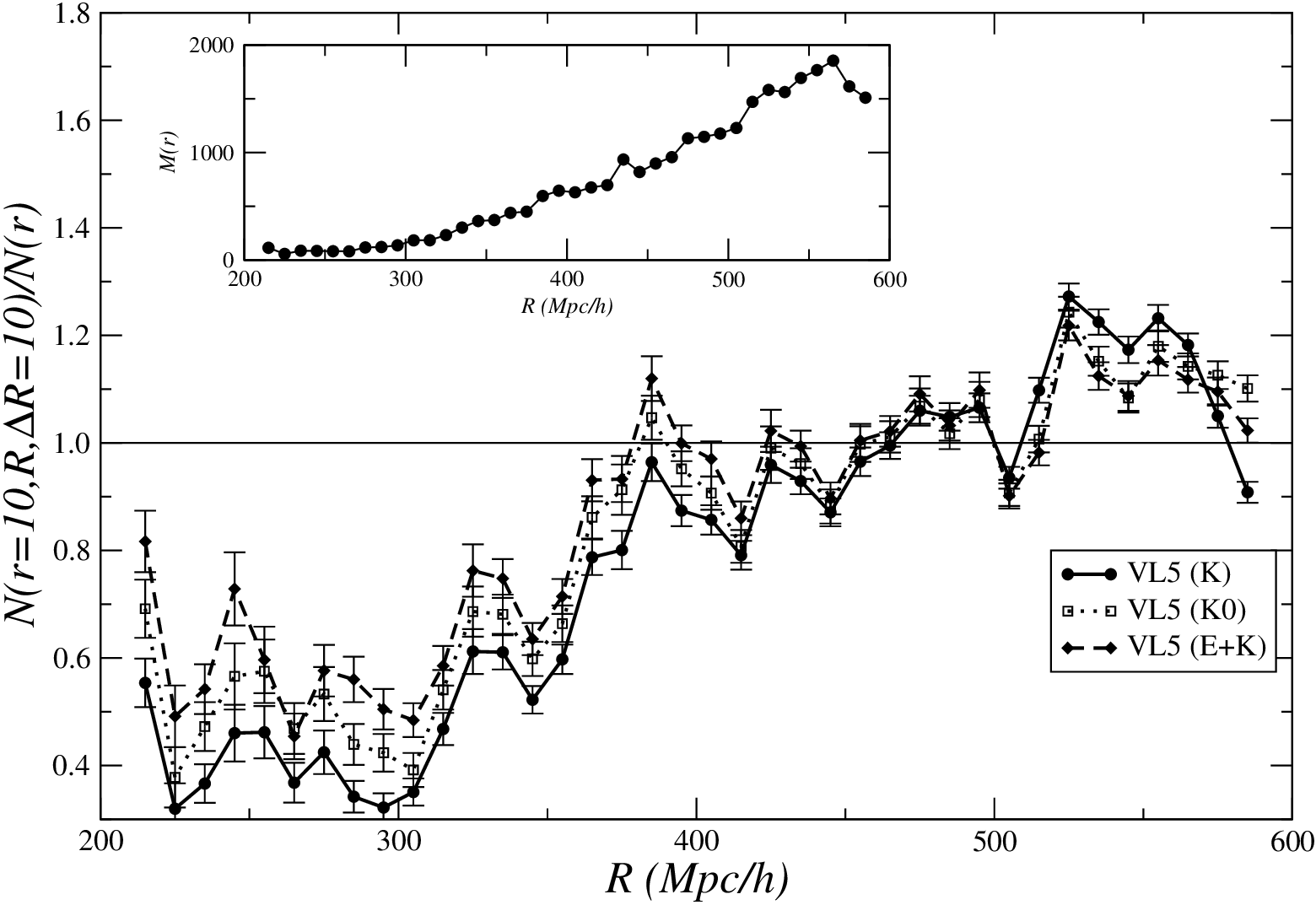}
\includegraphics*[angle=0, width=0.5\textwidth]{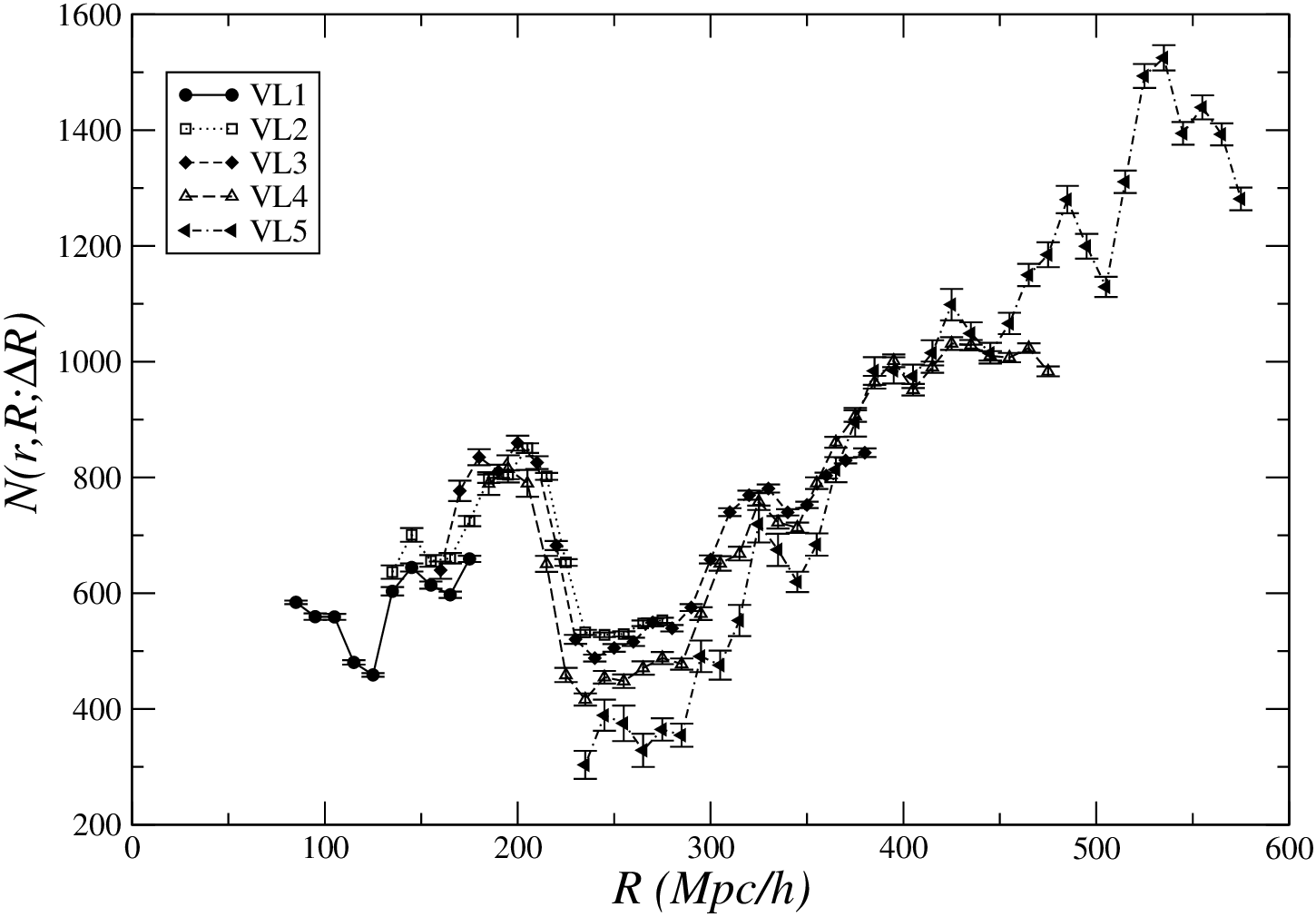}
\caption{Behavior of the local average of $\overline{N(r;R,\Delta R)}$
  (see Eqs.\ref{NRave}-\ref{NRvar}) normalized to the whole sample
  average (see Eq.\ref{eq1} below) in bins of thickness $\Delta R=10$
  Mpc/h for sphere radius $r=10$ Mpc/h normalized to the whole sample
  average for the 5 VL samples with K-correction (K), with evolution
  and K-correction (E+K), and without evolution and K-correction (K0).
  The insert panel shows the number of centers, over which the average
  and variance are computed in each $\Delta R$ bin.  In the bottom
  right panel we report the behavior of $\overline{N(r;R,\Delta R)}$
  in bins of thickness $\Delta R=10$ Mpc/h for $r=20$ Mpc/h,
  normalized to the luminosity factors as explained in the text (see
  Sect.\ref{norm:vl}), for K-corrected VL samples. }
\label{SL_VL1_K_E2K_K0}
\end{figure*}

To determine the features of galaxy structures on different scales, we
now consider a local average of $N_i(r)$ computed in the following
way. We divide the whole range of radial distances in each VL sample,
in bins of thickness $\Delta R$ and we compute the average
\be
\label{NRave}
\overline{N(r;R,\Delta R)}
 = \frac{1}{M_b} 
\sum_{R_j\in [R,\Delta R] }^{j=1,M_b} N(r; R_j)\;,
\ee
where the sum is extended to the $M_b$ determinations of
$N_i(r)=N(r;R_i)$ such that the radial distance of the $i^{th}$ center
is in the interval range $[R,R+ \Delta R]$. 
Its variance can be estimated by 
\bea
\label{NRvar}
&& 
\overline{\Sigma^2(r;R,\Delta R)}= 
\\ \nonumber 
&&
\frac{1}{M_b} 
\sum_{R_j\in [R,\Delta R] }^{j=1,M_b} 
\frac{
\left(\overline{N(r;R_j,\Delta R)} - N(r; R_j)\right)^2}
{(M_b-1)}
\;.
\eea

To study the sequence of structures and voids present in the samples,
we choose a relatively small radial bin, i.e. $\Delta R = 10$ Mpc/h,
and consider the sphere radius $r=10$ Mpc/h. It is clear that, as $r =
\Delta R$, there is some overlap in the determinations in contiguous
bins resulting in an artificial smoothing of the signal.  This means
that the fluctuations we detect in this way represent a {\it lower
  limit} to the real ones. In Fig.\ref{SL_VL1_K_E2K_K0} we show the
behavior of Eq.\ref{NRave} in bins of thickness $\Delta R=10$ Mpc/h
for sphere radius $r=10$ Mpc/h normalized to the whole sample average
(see Eq.\ref{eq1} below) for the three sets of five VL samples (region
R1) with different corrections, as defined in
Sect.\ref{sec:samples}. In VL1, VL2, and VL3 the signal is completely
unaffected by corrections while in VL4 and VL5 there is a small effect
that however, does not change the main trends.  In addition the insert
panels of Fig.\ref{SL_VL1_K_E2K_K0} shows the number of centers which
contribute to the average in each bin.  The fact that this grows as a
function of the radial distance reflects the limitations imposed by
the sample geometry discussed in Sect.\ref{sec:centers}.
Below we  summarize the situation.

\begin{itemize}

\item In the VL1 sample there are fluctuations of $\sim 40\%$. There
  is no a well-defined radial-distance trend; instead the scatter in
  the measurements corresponds to the location of large-scale
  structures.  The behavior is insensitive to the effect of the K
  and/or E corrections considered.

\item In the VL2 sample there is a high over-density in the radial
  distance range $[180,220]$ Mpc/h which is followed by a sharp decay,
  signaling a relative under-density for $R>220$ Mpc/h. Even in this case
  there is no detectable impact of K and/or E corrections
  considered.

\item The high over-density up to $R\approx 200$ Mpc/h is also visible
  in the VL3 sample, and is followed by an under-density in the range
  $220 < R < 270$ Mpc/h. Beyond 300 Mpc/h there is another relative
  over-density extending up to the sample boundaries.  The effect of
  E-corrections is to relatively amplify the over-density at $R\approx
  200$ Mpc/h with respect to the under-density on larger scales.

\item The behavior in VL4 is similar to the one in VL3.  Here the
  sharp fall in the average conditional density in bins at $\sim 220$
  Mpc/h is followed by a relatively slow growth, which seem to
  saturate at about $\sim 370$ Mpc/h at about the same level as the
  fluctuation at $\sim 200$ Mpc/h.  The effect of K and/or E
  corrections is to amplify the difference between amplitude of
  fluctuations at the short and long radial distances.

\item Even in the case of the sample VL5 the average behavior is
  quantitatively but not qualitatively changed by the effect of K
  and/or E corrections.  In this sample, as well as in VL4, there is a
  coherent trend over the whole sample volume, which is a signature of
  persisting large-scale fluctuations.
\end{itemize}


\subsection{Normalization of the behaviors in different VL samples} 
\label{norm:vl}

We can now normalize the behaviors of the radial density and of the
average conditional density in bins discussed in
Sect.\ref{sec:ave_bin} in the different VL samples.  This is done by
computing the normalizing factors for the different VL samples
assuming Eq.\ref{approxnu2} and by knowing the galaxy luminosity
function \citep{jsl01}.  In this approximation the observed radial
density in the VL1 sample can be written as
\be n^{VL1}(R)= n(R) \times
\int_{L_1}^{L_2} \phi(L) dL  = n(R) \times \Phi^{VL1} 
\ee 
where $L_1$ and $L_2$ are respectively the limit at the faint and
bright absolute luminosities of the sample VL1, and we have defined
\be \Phi^{VL1} = \int_{L_1}^{L_2} \phi(L) dL \;.  
\ee 
Clearly, the radial density, for instance, in the sample VL2, can be
normalized to that of VL1 by computing
\be 
\label{norma1} 
n^{VL1}(r) = \Phi^{VL1} \times \frac{n^{VL2}(R)} {\Phi^{VL2}} \;.
\ee 
Hereafter, to compute the normalization factors we use the best-fit
parameters to the luminosity function found in
Appendix.\ref{sec:lumfun}
\footnote{In Appendix \ref{sec:lumfun} we discuss the determination of
  the luminosity function and of two important, commonly used
  assumptions, that the space density is constant and that space and
  luminosity distributions are independent. We emphasize that the
  latter can be used also when the density field is inhomogeneous
  while the former corresponds to the strict assumption of spatial
  homogeneity.}.  The normalization factor for VL5 is the most
uncertain because the measured luminosity function deviates from the
simple Schechter function fit for bright magnitudes.

 Figure \ref{FIG_COUNTS_R1} shows the distance behavior of the
 normalized radial counts of galaxies in the region R1. A persistent
 growth of the density for distances $R>300$ is found, while for
 smaller radial distances there is the fluctuating behavior already
 discussed in the Sect.\ref{sec:ave_bin}. This is very
 similar~\footnote{although these behaviors look very similar, they
   refer to two different measurements which in principle are not
   expected to give the same behavior.}  to Fig.\ref{SL_VL1_K_E2K_K0}
 (bottom-right panel) where we considered the average in bins of the
 SL analysis, i.e. Eq.\ref{NRave}, as a function of the radial
 distance for $r=10$ Mpc/h with the same normalization factors as are
 used for the radial density. Indeed the same approximations as are
 used to derive the radial density normalization can be used to
 normalize the average SL data.  The normalization factors obtained in
 this way allow us to produce a single behavior from 50 Mpc/h to 600
 Mpc/h. The main features are again the over-density at $R\approx 200$
 Mpc/h, the relative low under-density in the range $[220,300]$ Mpc/h
 and the persistent growth for $R>300$ Mpc/h.

In this way we reach a completely different conclusion from that of
\citet{loveday}. Indeed, from the analysis of the luminosity function
for galaxies selected in four redshift slices ($0.001 < z < 0.1, 0.1 <
z < 0.15, 0.15 < z < 0.2$ and $0.2 < z < 0.3$) and despite the
uncertainties in the shape of the luminosity function in the redshift
slices, \citet{loveday} concluded that there is clear evolution in the
amplitude of the luminosity function, in the sense of an increasing
amplitude (vertical shift) and/or luminosity (horizontal shift) with
redshift.  On the other hand, we conclude that the behavior of the
radial counts of galaxies as a function of distance is consistent with
the average conditional number of galaxies in spheres as a function of
the radial distance i.e. Eq.\ref{NRave}.  The behaviors of
$\overline{N(r;R,\Delta R)}$ can be normalized simply by using the
results obtained in the same samples for the luminosity
function. Thus our conclusion is perfectly consistent with the
measurements of the PDF presented in the previous section, and it does
not imply that a strong evolution has occurred up to $z=0.2$. Rather,
as discussed above for the behavior of average conditional density in
bins, we can trace the various main structures in these samples:
namely there are large fluctuations at about 200 Mpc/h followed by a
large under-dense region up to 400 Mpc/h, which is then followed by
other coherent structures up to the sample limits.

\begin{figure}
\begin{center}
\includegraphics*[angle=0, width=0.5\textwidth]{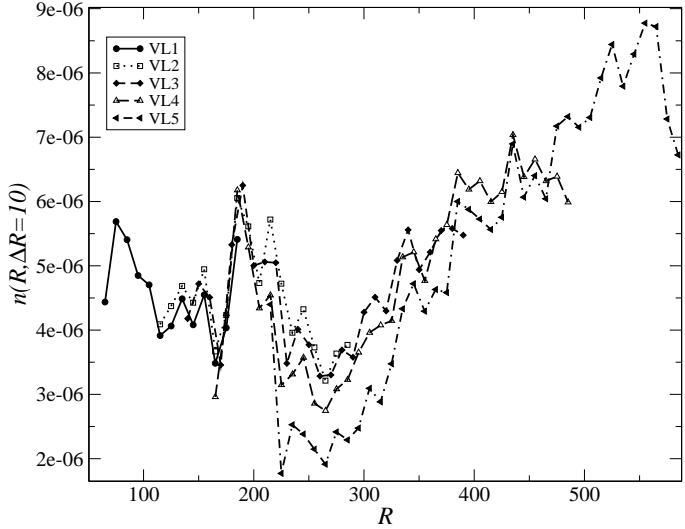}
\end{center}
\caption{Radial counts in bins of thickness $\Delta R=10$ Mpc/h,
 normalized to the luminosity factors as explained in the text, 
for the K-corrected VL samples.} 
\label{FIG_COUNTS_R1}
\end{figure}


\subsection{The whole sample average and the variance} 
\label{sec:gamma}

When Eq.\ref{eq1a} is averaged over the whole sample, it gives an
estimate of the average conditional density 
\be
\label{eq1} 
\overline{n(r)} = \frac{3}{4\pi r^3} \frac{1}{M(r)} 
\sum_{i=1}^{M(r)} N_i(r) \;.
\ee
In Fig.\ref{Gamma_R1} we show the whole-sample average conditional
density in the different K-corrected VL samples, normalized by using
Eq.\ref{norma1}. Contrary to the behavior of the radial number counts
and of the SL statistics averaged in bins (Eq.\ref{NRave}), in this
case the behavior of the average conditional density in different
samples do not overlap in a satisfactory way.  This is due to the fact
that the whole-sample average is biased by the lack of self-averaging
properties and it does not give a reliable estimation of the ensemble
quantity. Regardless of its amplitude the quantity $\overline{N(r)}$
shows a power law behavior with exponent $D=2.2 \pm 0.1$ up to $\sim
30$ Mpc/h. On larger scales, its determination is strongly affected by
the non-self averaging properties of conditional fluctuations
discussed above. To reliably detect uniformity, the conditional
density has to be flat for a wide enough range of scales, while in the
data we measure a different scale-dependence for $r>20$ Mpc/h than on
small scale, but we cannot detect a clear flattening.  However, in
view of the large fluctuations detected by the complete PDF analysis
and by the self-averaging test, we conclude that there is no crossover
to uniformity up to $\sim$ 100 Mpc/h.  We need to consider larger
samples to properly constrain correlations properties for scales
greater than $r>30$ Mpc/h.

\begin{figure}
\begin{center}
\includegraphics*[angle=0, width=0.5\textwidth]{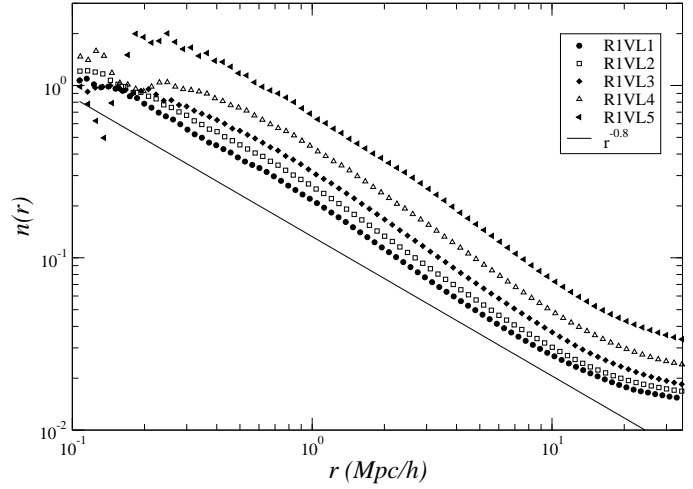}
\end{center}
\caption{Whole-sample average conditional density in the different
  K-corrected VL samples in the regions R1, normalized as explained in
  the text. }
\label{Gamma_R1}
\end{figure}

As mentioned above, for $r<30$ Mpc/h in all samples the PDF is stable
with respect to the K+E-corrections, so there are no detectable
differences in the three sets of VL samples with different
corrections.  Thus, while for scales $r<30$ Mpc/h the data show an
approximated power law behavior for $r>30$ Mpc/h, we are not able to
make a reliable conclusion, in these samples, about the behavior of
this quantity, because conditional fluctuations do not exhibit
self-averaging properties when filtered on scales $r>30$ Mpc/h. The
large-scale inhomogeneity shown by the non self-averaging conditional
fluctuations is compatible with a continuation of power-law
correlations, i.e. scaling properties, to scales larger than 30 Mpc/h.


\subsection{The standard two-point correlation function}

When determining the standard two-point correlation function, we
implicitly make two assumptions that, inside a given sample, (i) the
distribution is self-averaging and (ii) it is uniform.  The first
assumption is used when computing whole-sample average quantities. For
instance it is assumed when the whole-sample average conditional
density is measured, as discussed in the previous section. However, in
that case, there is need to assume that the estimation of the sample
average gives a fairly good estimation of the ensemble average
density. This corresponds to the assumption (ii) above. When one of
these assumptions, or both, is not verified then interpretation of the
results given by the determinations of the standard two-point
correlation function must be reconsidered with great care as we
discuss in what follows.

To measure the two-point correlation function, the most commonly used
estimators are based on pair-counting algorithms, as the Davis and
Peebles \citet{dp83} (DP) and the \citet{lz} (LS) estimators. These
are relatively easy to implement practically by generating random
distributions in artificial samples with the same geometry as real
ones.  In general, it is not straightforward to interpret the results
obtained with these estimators at large enough scales; i.e., from
around the scale $r^{ps}$ at which the spherical shell of radius
$r^{ps}$, centered on a typical distribution point, is only partially
contained in the sample volume ~\citep[see][405]{book}. The scale
$r^{ps}$ is  the one up to which one can calculate the so-called
full-shell (FS) estimator, i.e. in which only complete spherical
shells are considered ~\citep{kerscher}.

The FS estimator considers, similarly to the case of the conditional
density estimator, a pair of points at distance $r$ only if a sphere
of radius $r$, centered on one of the points, is fully contained in
the sample volume. Thus this method, because it requires fewer
assumptions, is the one we consider in more detail here.  The FS
estimator can be written as \citep{book}
\be
\label{xi} 
\xi(r) +1 = \frac{\overline{N(r,\Delta r)}} {V(r,\Delta r)} \cdot
\frac{1} {n_s} \;.  
\ee 
The first ratio in the r.h.s. of Eq.\ref{xi} is the average
conditional density, i.e., the number of galaxies in shells of
thickness $\Delta r$ averaged over the whole-sample, divided by the
volume $V(r,\Delta r)$ of the shell. The second ratio in the r.h.s. of
Eq.\ref{xi} is the density estimated in a sample containing $N$
galaxies, with volume $V$.  Thus, the FS estimator requires
determination of the distances of all points to the boundaries as for
the case of the conditional density (see Eq.\ref{eq1a} and
Eq.\ref{eq1}).  However, it should be stressed that, when measuring
this function, we implicitly assume in a given sample, that (i)
fluctuations are self-averaging in different subvolumes and (ii) the
linear dimension of the sample volume is $V^{1/3} \gg \lambda_0$
\citep{book}, i.e., the distribution has reached homogeneity inside
the sample volume.  If one of them, or both, is not verified in the
actual data, then the amplitude and shape of the estimated $\xi(r)$
will strongly depend on the sample volume. This finite-size dependence
can be investigated by making specific tests as we discuss in what
follows. We stress that the most efficient way to test the above
assumptions is represented by the determination of the conditional
fluctuations presented in the previous sections.

To show how non self-averaging fluctuations inside a given sample bias
the $\xi(r)$ analysis, we consider the estimator
\be 
\label{xi2}
\xi(r;R,\Delta R) +1 = \frac{\overline{N(r,\Delta r)}} {V(r, \Delta
  r)} \cdot \frac{V(r^*)}{\overline{N(r^*;R,\Delta R)}} \,, 
\ee 
where the second ratio on the r.h.s. is now the density of points in
spheres of radius $r^*$ averaged over the galaxies lying in a shell of
thickness $\Delta R$ around the radial distance $R$.  If the
distribution is homogeneous, i.e., $r^*>\lambda_0$, and statistically
stationary, Eq.\ref{xi2} should be statistically independent on the
range of radial distances $(R,\Delta R)$ considered.

Indeed the two-point correlation function is defined as a ratio
between the local conditional density and the sample average density:
if both vary in the same way when the radial distance is changed, then
its amplitude remains nearly constant. This does not imply, however,
that the amplitude of $\xi(r)$ is meaningful as the density estimated
in subvolumes of size $r^*$ can show large fluctuations, and this
occurs with a radial-distance dependence. To show that the $\xi(r)$
analysis gives a meaningful estimate of the amplitude of fluctuations,
{\it one has to test that this amplitude remains stable by changing
  the relative position of the subvolumes of size $r^*$ used to
  estimate the local conditional density and the sample average
  density}. This is achieved by using the estimator in
Eq.\ref{xi2}. On the other hand, standard estimators are unable to
test for such an effect, as the main contributions for both the local
conditional density and the sample average density come from the same
part of the sample (typically the more distant part where the volume
is larger).

For instance we consider, in the VL3 sample, $\Delta R=50$ Mpc/h and
$R = 250$ Mpc/h or $R=350$ Mpc/h, with $r^* = 60$ Mpc/h. We thus find
large variations in the amplitude of $\xi(r)$ (see
Figs.\ref{figxiVL3}-\ref{figxiVL5}).
\begin{figure}
\begin{center}
\includegraphics*[angle=0, width=0.5\textwidth]{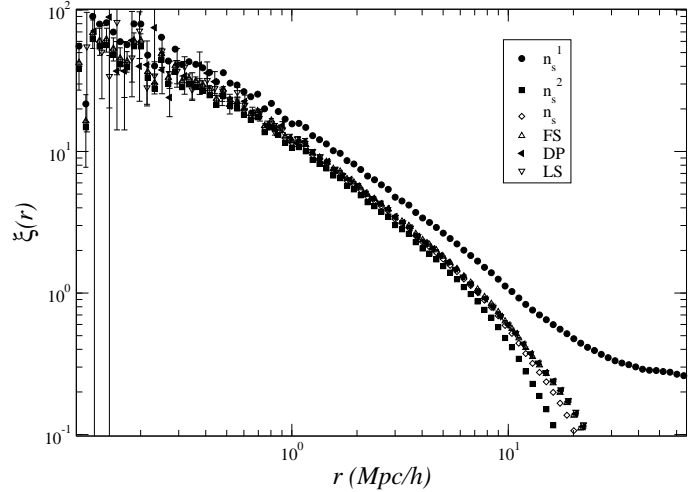}
\end{center}
\caption{Standard two-point correlation function in the VL3 sample
  estimated by Eq.\ref{xi2}: the sample average density is computed in
  spheres of radius $r^*=60$ Mpc/h and considering all center-points
  lying in a bin of thickness $\Delta R=50$ Mpc/h centered at
  different radial distance $R$: $R_1=250$ Mpc/h ($n_s^1$) and
  $R_2=350$ Mpc/h ($n_s^2$).  The case in which we have used the
  estimation of the sample average $N/V$ ($n_s$) is also shown and it
  agrees with the FS estimator. This former agrees with the
  measurements provided by the LS and DP estimators which give
  essentially the same result. (For sake of clarity error bars are
  shown for the FS, DP and LS estimators, and they are relatively
  small except at small and large $r$).}
\label{figxiVL3}
\end{figure}
\begin{figure}
\begin{center}
\includegraphics*[angle=0, width=0.5\textwidth]{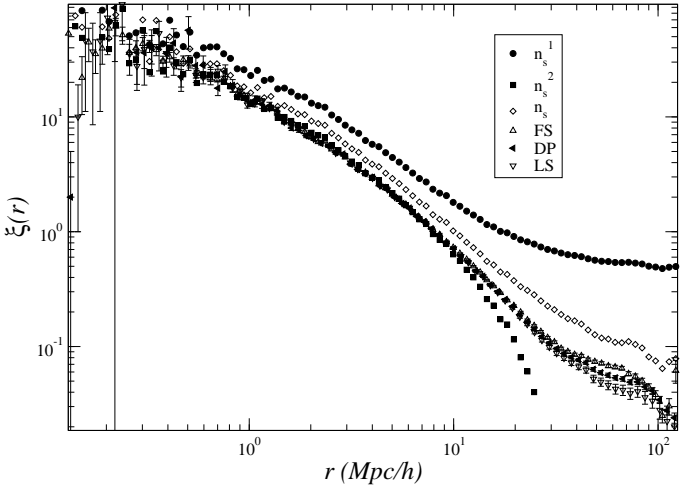}
\end{center}
\caption{The same as in Fig.\ref{figxiVL3} but now for the VL5
  sample. In this case the sample average density is computed in
  spheres of radius $r^*=80$ Mpc/h and considering all center-points
  lying in a bin of thickness $\Delta R=80$ Mpc/h centered at
  different radial distance $R$: $R_1=320$ Mpc/h ($n_s^1$) and
  $R_2=450$ Mpc/h ($n_s^2$).  }
\label{figxiVL5}
\end{figure}
This is simply an artifact generated by the large density fluctuations
on scales close to the sample sizes.  The results that the estimator
Eq.\ref{xi}, or others based on pair counting \citep{book,cdm_theo},
has nearly the same amplitude in different samples, e.g.,
\citep{dp83,park,benoist,zehavi_earlydata,zehavietal05,norbergxi01,norbergxi02},
despite the large fluctuations of $N_i(r;R)$, are simply explained by
the fact that $\xi(r)$ is a ratio between the local conditional
density and the sample average density. Both vary in the same way when
the radial distance is changed, so the amplitude is nearly constant.

To understand how large-scale fluctuations can be hidden in the
analysis performed by the two-point correlation function we consider
the following simple example. Let us suppose the catalog consists of
two disconnected volumes: for simplicity we fix them to be spherical
with radii $R_s^1$ and $R_s^2$ respectively. Let us suppose that the
average conditional density (supposed to be self-averaging at all
scales considered) is power-law, i.e.
\be 
\overline{n^{1,2}(r)} = \frac{\overline{N(r,\Delta r)}}
{V(r, \Delta r)} = \frac{(3-\gamma) B^{1,2}}{4 \pi} r^{-\gamma} \;,
\ee 
where $B^1$
is the amplitude in the volume $V^1$ and $B^2\ne B^1$ in the volume
$V^2$.  The estimation of the sample density is 
\be 
n_S^{1,2} = \frac{3}{4 \pi (R_s^{1,2})^3}  \int_{V(R_s)} 
\overline{n^{1,2}(r)} d^3r = 
 \frac{N}{V} = \frac{3 B^{1,2}}{4 \pi (R_s^{1,2})^\gamma} \;.
\ee
It is clear that if $B^2\ne B^1$ there will be large fluctuations
between the two volumes on scales close to the sample sizes.  However
if $R_s^1 = R_s^2=R_s$, from Eq.\ref{xi} we find that the estimator of
the two-point correlation function is
\be 
\label{eq23}
\overline{\xi_{1,2}(r)} =
\frac{\overline{n^{1,2}(r)}}{n_S^{1,2}}-1= \frac{3-\gamma}{3} \left(
\frac{r}{R_s^{1,2}} \right)^{-\gamma} -1 \;.  
\ee
This no longer depend on the different amplitudes of the conditional
density.  That is, despite the difference in the conditional density
and in the whole-sample density in the two volumes (which depends on
the ratio between $B^1$ and $B^2$), the amplitude of the two-point
correlation function does not reflect these (arbitrarily large)
variations.

Similarly in the case $R_s^1 \ne R_s^2$, the difference in amplitude
between the estimation of the two-point correlation function in the
two volumes is simply
\be 
\label{diffamp}
\frac{\overline{\xi_{1}(r)+1}}{\overline{\xi_{2}(r)+1}} =
\left(\frac{R_s^2}{R_s^1}\right)^{-\gamma} \;, 
\ee
thus resulting in a relatively small factor, when $R_s^1 \approx
R_s^2$ \footnote{For a sample of arbitrary geometry $R_s$ is defined
  to the  radius of the largest sphere fully contained in the
  sample volume \citep{book}.}, even though the difference between
$B^1$ and $B^2$ can be arbitrarily large!  That is, even though the
average density can fluctuate by an arbitrarily large factor, the
amplitude of $\xi(r)$ may not show a similar variation. This does not
imply, however, that the amplitude measures an intrinsic property of
the distribution.  Actually, in Eq.\ref{diffamp} the difference in the
amplitude is related to the sample sizes. This means that the only
unambiguous way to establish whether the average density is a
well-defined quantity,  hence whether the results obtained by the
standard correlation function analysis are meaningful, is represented
by the study of conditional fluctuations presented in the previous
sections.

By using different normalizations, which however are all in principle
equally valid if the distribution has a well-defined average density
inside the sample, we have shown that the amplitude of the estimated
correlation function varies in the SDSS samples. This occurs because
both the assumptions on which the determination of the standard to
point correlation function is based, are not verified in these
samples, and  $\lambda_0$ is certainly larger than the samples
size.

Finally we note that, not only the amplitude, but also the shape of
the correlation function is affected by the normalization to a sample
average, which largely differs from the ensemble average one. The
shape however is strongly biased only on large separations when
$\xi(r) \ll 1$, i.e. when the first term in the r.h.s. of
Eq.\ref{eq23} becomes comparable to the second one. We refer the
interested reader to \citet{martin} for a more detailed discussion of
the determination of the standard estimators (i.e., the LS and DP
estimators) of the two-point correlation function in these samples.


\subsection{The SDSS Great Wall and other structures}

\begin{figure*}
\centering
\includegraphics*[width=17cm]{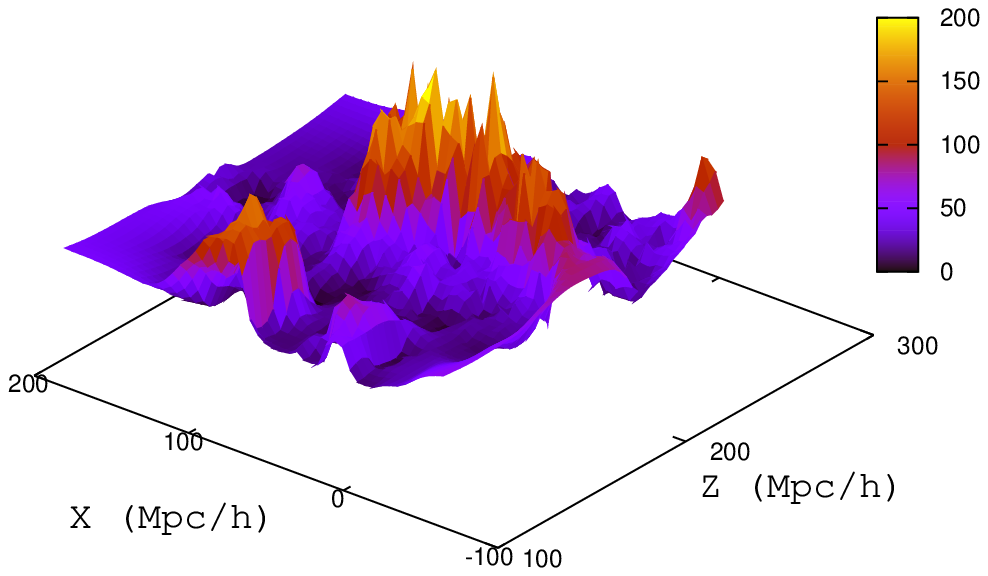}
\caption{Three-dimensional representation of the SL analysis with
  $r=10$ Mpc/h for R3VL2. The $x,z$ coordinates of the sphere center
  define the bottom plane,and on the vertical axis we display the
  intensity of the structures, the conditional number of galaxies
  $N_i(r)$ contained in the sphere of radius $r$.}
\label{SDSSGw_col}
\end{figure*}

As mentioned above, the measurements of the $M(r)$ values of $N_i(r)$
on the scale $r$ allow derivation of many interesting properties about
structures in these samples. Beyond the statistical properties already
described, it is interesting for example to consider the density
profile derived from $N(r;R_i)$. An example is shown in
Fig.\ref{SLVL2-R1R2R3}, which displays the behavior of $N(r;R_i)$ in
the sample VL2 (with K-corrections) and in the three different regions
for $r=10$ Mpc/h.
\begin{figure}
\begin{center}
\includegraphics*[angle=0, width=0.5\textwidth]{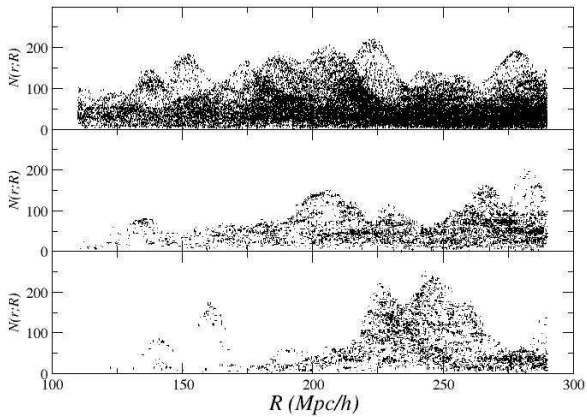}
\end{center}
\caption{Behavior of $N(r;R_i)$ in the K-corrected VL2 sample and in
  the three different regions for $r=10$ Mpc/h (R1 top, R2 Middle and
  R3 bottom).}
\label{SLVL2-R1R2R3}
\end{figure}
This analysis is more powerful than the simple counting as a function
of radial distance, in tracing large-scale galaxy structures. Indeed,
one may precisely describe the sequence of structures and voids
characterizing the samples and, by changing the sphere radius $r$, one
may determine the situation at different spatial resolutions.  For
instance, the distribution in the angular region R3 (see the bottom
panel of Fig.\ref{SLVL2-R1R2R3}) is dominated by a single large-scale
structure, which is known as the SDSS Great Wall \citep{gott}. In the
R2 and R3 regions, one is also able to isolate structures well at
different distances, while the R1 region, which covers a solid angle
about six times larger than the other two sky areas, the signal is
determined by the superposition of different structures of different
amplitude and on different scales. In the latter case, it would be
useful to divide the sample into smaller angular slices.

In Fig.\ref{SDSSGw} we show the projection on the $X-Z$ plane of R3VL2
where the SDSS Great Wall is placed in the middle of the sample, and
it is clearly visible as a coherent structure of large amplitude ,
similar to a mountain chain, extending over the whole sample. The
information contained in the $N(r;R_i)$ data allow quantitative
determination of the properties of this structure in an unambiguous
way, as we discussed above.  For instance by a simple visual
comparison of the profile in the different angular region we can
conclude that, although the Great Wall is a particularly long filament
of galaxies, it represents a typical persistent fluctuation in the
samples' volume.
\begin{figure}
\begin{center}
\includegraphics*[angle=0, width=0.5\textwidth]{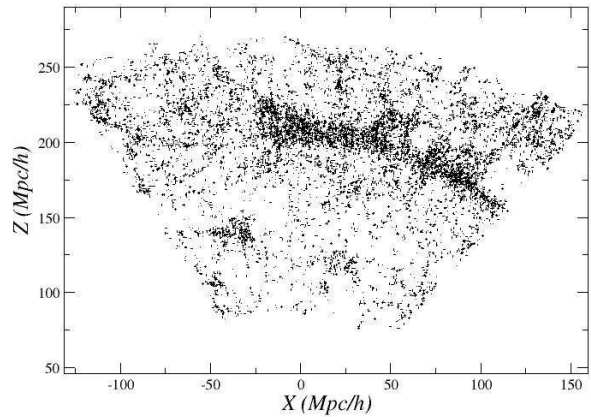} 
\end{center}
\caption{Projection on the $X-Z$ plane of R3VL2. The SDSS Great Wall
  is the filament in the middle of the sample.}
\label{SDSSGw}
\end{figure}

In addition it is interesting to consider the full $N_i(r) = N(r;
x_i,y_i,z_i)$ data, where $(x_i,y_i,z_i)$ are the Cartesian
coordinates of the $i^{th}$ center. To this aim we chose a
three-dimensional representation where on the bottom plane we use the
$x,z$ Cartesian coordinates of the sphere center and on the vertical
axis we display the intensity of the structures, i.e.,  the conditional
number of galaxies contained in the sphere of radius $r$.  (In the $y$
direction the thickness of the sample is small, i.e., $\Delta y \approx
15$ Mpc/h.)  This is shown in Fig.\ref{SDSSGw_col} for $r=10$
Mpc/h. One may note that that the SDSS Great Wall is clearly visible
as a coherent structure similar to a mountain chain, extending all
over the sample.  It is worth noticing that profiles similar to those
shown in Figs.\ref{SDSSGw_col}-\ref{SLVL2-R1R2R3} have also been
found in the 2dFGRS \citep{paper_2df_prl,paper_2df_aea}
supporting  that the fluctuations we have identified in this
catalog are  typical of galaxy distribution.


\subsection{Role of spatial correlations} 

To show that the large-scale fluctuations in the galaxy density field
we have detected are genuinely due to long-range spatial correlations
and not to some selection effects, we performed the following test.
In a given VL sample we have assigned a redshift randomly to each
galaxy extracted from the list of redshifts of the galaxies in the
same sample~\footnote{We are grateful to David Hogg for interesting
  suggestions about this test.}. In this way the angular coordinates
of each object are fixed, its redshift is randomized while the
redshift distribution in the sample is taken fixed.  This operation
washes out the intrinsic spatial correlations of the galaxy
distribution, but conserves the main observational coordinates
(i.e.. angular positions and redshift). Thus the result of this test
may tell us whether fluctuations and structures are an effect of
spatial correlations.  The results is that the signal in $N_i(r,R)$ is
substantially washed out as one may noticed by comparing
Figs.\ref{SLVL2-RND-R1R2R3}-\ref{SLVL2-R1R2R3}.
\begin{figure}
\begin{center}
\includegraphics*[angle=0, width=0.5\textwidth]{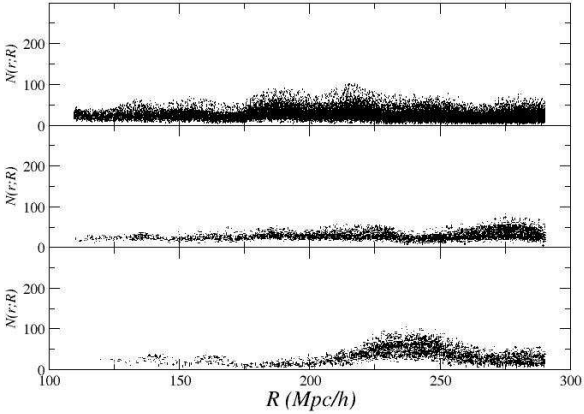}
\end{center}
\caption{As in Fig.\ref{SLVL2-R1R2R3} but for  the randomized VL2 samples
as described in the text.} 
\label{SLVL2-RND-R1R2R3}
\end{figure}

However, it should be stressed that, if there are structures of
spatial extension comparable to the sample size, these will not be
completely washed out by the randomization adopted given that the
redshift distribution is taken to be fixed.  Indeed this is the case
for the SDSS Great Wall, contained in the sample R3VL2.  In
Fig.\ref{SLVL2-RND-R1R2R3} the structure is almost completely washed
out, but as it is as large as the sample, there is a residual in the
randomized version.  By means of the the statistical analysis shown in
Fig.\ref{R3VL2_RND}, we find that the PDF of conditional fluctuations
becomes very peaked in the randomized sample; i.e., it tends to a
Gaussian function, while in the real sample it displays a long tail
for high $N$ values, corresponding as discussed above to the large
fluctuations present in this sample.  In addition the conditional
density (i.e., the conditional average number of points in spheres
given by Eq.\ref{eq1} divided by the spherical volume of radius $r$)
becomes flat for the randomized sample, signaling the absence of
correlations, while it was a power law in the real sample with an
exponent approximately equal to $0.8 \pm 0.1$.  Because the redshift
distribution is taken as fixed in the randomized sample, its PDF will
converge to the one of the real sample by considering larger sphere
radii.
\begin{figure}
\begin{center}
\includegraphics*[angle=0, width=0.5\textwidth]{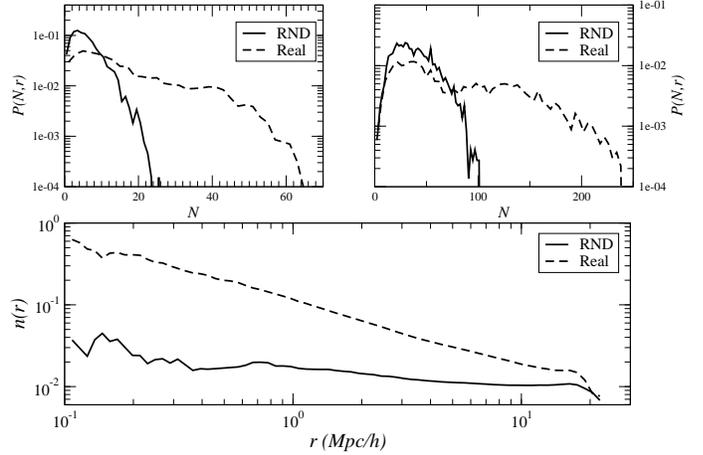}
\end{center}
\caption{{\it Upper panels}: the PDF of conditional fluctuations in
  spheres of radius $r=5$ Mpc/h (left) and $r=10$ Mpc/h (right) for
  the real sample (Real) and the randomized one (RND) as explained in
  the text. {\it Bottom panel}: conditional density as a function of
  scale. }
\label{R3VL2_RND}
\end{figure}


\section{Comparison with  theoretical models}
\label{sec:mock}

Let us now discuss the problem of comparing the statistical analysis
of real galaxy samples with theoretical predictions. In this respect
it is useful to remember that theoretical models predict that, by
gravitationally evolving a density field compatible with cosmic
microwave background anisotropies (CMBR) observations, there is a
maximum scale up to which nonlinear structures have formed at the
present time. The precise value of such a scale depends on the details
of the initial correlations in the density field and on the values of
the cosmological parameters, and this is roughly placed at about
$\lambda_0\approx 10$ Mpc/h in the $\Lambda$CDM concordance model 
\citep[see][]{springel05}.  On scales of $r>\lambda_0$ the average
density becomes well-defined as long its fluctuations become small
enough. As discussed in Sect.2, this scale may be defined as the one
at which the variance of the fluctuations is twice the square of the
asymptotic (large-scale) average density. Then for scales
$r>\lambda_0$, the situation is simple: gravitational clustering has
linearly amplified initial fluctuations and thus correlation
properties reflect those at the initial time. The linear amplification
factor can be easily computed by making a perturbation analysis of the
self-gravitating fluid equations (i.e., Vlasov-Poisson equations) in an
expanding universe.

There is no full analytical understanding of the properties of
self-gravitating particles in the nonlinear phase occurring for scales
$r<\lambda_0$; for this reason, generally gravitational N-body
simulations represent the means to study these structures. A
gravitational N-body simulation follows the motion of particles
(supposed in cosmology to be dark matter particles) moving under the
effect of their self-gravity in an expanding universe. By normalizing
the initial amplitude of fluctuations and the density correlations to
the observations of the CMBR anisotropies, one finds that there is a
well-defined time scale that allows one to define the present time at
which the simulation is stopped.  In this context it is worth
remembering that, in the CDM-like models gravitational clustering
builds up non-linear structures in a bottom-up way because of the
small initial velocity dispersion.

An N-body simulation provides the distribution of dark matter
particles and not that of galaxies. It is here that the problem of
sampling, or biasing, is relevant. Indeed galaxies are supposed to
form on the highest peaks of the dark matter density field, so one has
to define the rules to make a {\it correlated} sampling of the dark
matter particles to identify ``mock'' galaxies. There are different
sampling procedures in the literature and they are the outcome of the
so-called semi-analytic models of galaxy formation.  Generally these
sampling procedures only modify correlations at small scales, i.e.,
they are {\it local sampling}. Only nonlocal sampling procedures may
give rise to different correlation properties on large scales.
However, no form of currently known galaxy bias can produce the
large-scale fluctuations we observe in the catalog.  Indeed the
current accepted theoretical model of biasing \citep{kaiser} predicts
that, when clustering is in the linear phase, threshold sampling the
highest peaks in a Gaussian density field gives rise to a simple
linear amplification of fluctuations and of their correlations.  This
situation is expected to hold for scales $r>10$ Mpc/h where density
fluctuations in N-body simulations of the dark matter field are in the
linear regime, the PDF of fluctuations is Gaussian, and thus biasing
is linear. In these conditions, there is a simple relation between
mock galaxies and dark matter particle correlation properties. Thus
complications with respect to this simple picture are expected only on
small scales.

We use a semi-analytic galaxy catalog constructed from the Millennium
$\Lambda$CDM N-body simulation \citep{springel05}. To construct mock
samples corresponding to SDSS VL samples, we used full version of the
catalog in the $ugriz$ filter system. The catalog contains about 9
million galaxies in a 500 Mpc/h cube \footnote{see {\tt
    http://www.mpa-garching.mpg.de/galform/agnpaper/} for
  semi-analytic galaxy data files and description, and see {\tt
    http://www.mpa-garching.mpg.de/millennium/} for information on
  Millennium LCDM N-body simulation.}.  We used the absolute
magnitudes in $r$ filter used in the SDSS case, to construct the mock
samples with the same limits in absolute magnitude as for the SDSS VL
samples with K-corrections. In Table~\ref{tbl_VLSamplesPropertiesmock}
we report the properties of the mock samples: $R_{min}$, $R_{max}$ (in
Mpc/h) are the chosen limits for the metric distance; ${M_{min},
  \,M_{max}}$ define the interval for the absolute magnitude in each
sample, $\alpha_{max},\alpha_{min}$ (in degrees) the limits in right
ascension, $\delta_{max},\delta_{min}$ (in degrees) the limits in
declination, $N_z$ the number of objects in the sample in
redshift-space and $N_r$ the same for the sample in
real-space~\footnote{The difference between real and redshift-space is
  due to peculiar velocities.}. In addition, we construct only the mock
samples corresponding to VL1, VL3 and VL5. The volume of the samples
is constrained to be the same as, or similar to, the volumes of the
real SDSS samples. In particular, the sample region can be easily
fitted in the simulation cube in case of VL1. For the VL3 it should be
slightly reduced in declination, while is reduced significantly for
the VL5 the range in declination (see
Table~\ref{tbl_VLSamplesPropertiesmock}).

\begin{table*}
\begin{center}
\begin{tabular}{|c|c|c|c|c|c|c|c|c|}
  \hline Sample & $R_{min}$ (Mpc/h) & $R_{max}$ (Mpc/h) &
  $\alpha_{min}$ & $\alpha_{max}$ & $\delta_{min}$ & $\delta_{max}$ &
  $N_r$ & $N_z$ \\ \hline VL1 & 50 & 200 & 24.0 & 66.0 & -48.0 & 32.5
  & 53423 & 54555 \\ VL3 & 125 & 400 & 24.0 & 66.0 & -45.0 & 30.0 &
  74645 & 74170 \\ VL5 & 200 & 600 & 24.0 & 66.0 & -24.5 & 24.5 &
  15572 & 15571 \\

   \hline
\end{tabular}
\end{center}
\caption{Main properties of the obtained mock VL samples.}
\label{tbl_VLSamplesPropertiesmock}
\end{table*}

The PDF of conditional fluctuations (see Fig.\ref{PDF_VL1_mock}) show
a clear departure from a Gaussian function for $r<10$ Mpc/h, while it
rapidly approaches the Gaussian function for $r>20$ Mpc/h. For $r>5$
Mpc/h there is no detectable difference between the real and
redshift-space cases.  Additionally, for $r<10$ Mpc/h, the PDF
exhibits a large $N$ tail that is the signature of the correlations
present at those scales.

\begin{figure*}
\includegraphics*[width=17cm]{Fig20.eps}
\caption{{  PDF of conditional fluctuations in the mock R1VL1, R1VL3,
    and R1VL5 samples in real (red line) and redshift-space (black
    line). (Each row corresponds to a VL sample; the scale $r$ is
    reported in the caption). The best-fit Gaussian function (green
    line) is reported.}}
\label{PDF_VL1_mock}
\end{figure*}

\begin{figure}
\begin{center}
\includegraphics*[angle=0, width=0.5\textwidth]{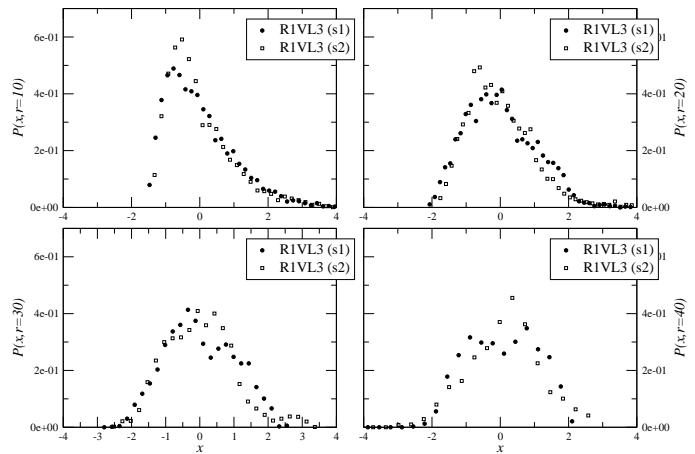}
\end{center}
\caption{The self-averaging test for the mock catalogs. It is
    analysis of the PDF into two disconnected subvolumes ($s_1$ and
    $s_2$) of the mock sample R1VL3 in redshift-space. }
\label{PDF_VL1_mock2}
\end{figure}

{\it In addition the analysis of the PDF into two disconnected
  subvolumes of the mock samples does not show any trend forward non
  self-averaging and they coincide in a statistical sense (see
  Fig.\ref{PDF_VL1_mock2})}. That is, in contrast to the case of real
galaxy samples, the simulations are self-averaging.

Figure \ref{Gamma_mock} shows the behavior of the conditional average
density in the mock samples. The main difference between real and
redshift-space occurs at scales $r<5$ Mpc/h, where the redshift-space
exponent is systematically smaller than the real space one.  For the
case of real galaxy samples we cannot make the real-space analysis
because galaxy peculiar velocities are not known. We noticed in that
the same finite-size effects that perturb the redshift-space reduced
two-point correlation function may affect the projected one, and thus
the whole method to infer the real-space correlation function from the
redshift-space one \citep{2df_paper}.  In addition, both in real and
redshift-space, the exponent is smaller when the average galaxy
luminosity increases, a trend that is not as well-defined in the real
data as shown by Fig.\ref{Gamma_R1}.  In the mock catalogs, the
power-law behavior extends up to $\sim 20$ Mpc/h, beyond which there
is a well-defined crossover which corresponds to the scale where PDF
of conditional fluctuations approaches the Gaussian function.  Thus,
while the exponent of the conditional density is closer in
redshift-space, up to $r \sim 10$ Mpc/h, to what is observed in the
real galaxy data, the mock samples show a clear difference for $r>20$
Mpc/h, in that the crossover to homogeneity is well-defined and the
distribution does not present large-scale fluctuations similar to
those characterizing the SDSS galaxy distribution. In addition,
redshift and real-space properties are indistinguishable for $r>10$
Mpc/h.

\begin{figure}
\begin{center}
\includegraphics*[angle=0, width=0.5\textwidth]{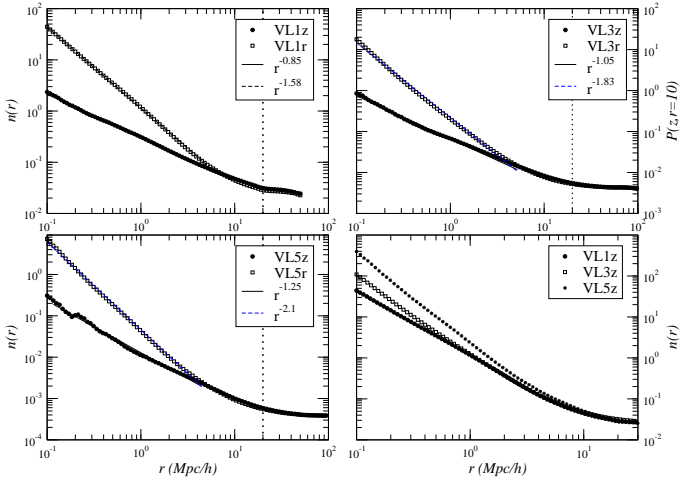}
\end{center}
\caption{Conditional density in the mock R1VL1, R1VL3, and R1VL5 sample
  in real (e.g., VL1r, VL3r, VL5r) and redshift-space (e.g., VL1z, VL3z,
  VL5z). In the panel on the bottom left there is a comparison of the
  behaviors in the different samples in redshift-space, where it is
  evident that the exponent becomes steeper for brighter objects. The
  normalization is taken to have the same large-scale density.}
\label{Gamma_mock}
\end{figure}


As a final remark, we note that \citet{einasto1} find that the
fraction of very luminous (massive) superclusters in real samples
extracted from 2dFGRS and from the SDSS (Data Release 4) is more than
ten times greater than in simulated samples constructed from the
Millennium simulations --- see also \citet{einasto2,einasto3}. Our
results are compatible with these findings.


\section{Conclusion}
\label{sec:discussion}

We have studied the statistical properties of galaxy distribution from
the SDSS-DR6 sample.  This is brief summary of our results:
\begin{itemize}

\item The probability density function (PDF) of spatial conditional
  fluctuations, in volume limited samples and filtered on spatial
  scales $r<30$ Mpc/h, shows a long tail, which is the signature of
  the large-scale structures in these samples.
\item The PDF of conditional fluctuations does not converge to a
  Gaussian function even for large-sphere radii (i.e., $r>$ 30 Mpc/h).
\item The PDF of conditional fluctuations is unaffected by K and
  (standard) evolutionary corrections.
\item The PDF of conditional fluctuations, filtered on spatial scales
  $r>30$ Mpc/h, does not show self-averaging properties when this is
  computed in two non-overlapping samples of equal volume.
\item The whole-sample averaged conditional density shows scaling
  properties up to $\sim 30$ Mpc/h, the largest scales where this
  statistics shows self-averaging properties (i.e, where the PDF is
  statistically stable inside the sample).
\item The normalization of the luminosity-redshift function and of the
  two-point correlation function are affected by systematic
  effects,  thus do not provide  meaningful information.
\item The PDF of conditional fluctuations in mock galaxy catalogs
  rapidly converges to a Gaussian function for $r>10$ Mpc/h so that
  structures predicted by theoretical models are at odds with
  observations.
\item The above behaviors are compatible with galaxy counts as a
  function of apparent magnitude and of redshift in the
  magnitude-limited sample. Indeed these show fluctuations on large
  spatial scales of $\sim 10 \div 20 \%$, which are persistent in the
  sample volume.
\end{itemize}

In summary, the primary conclusion of this work is that in the
SDSS-DR6 we find large-scale galaxy structures that correspond to
density fluctuations of large amplitude and large spatial extension,
whose size is only limited by the sample size.  Because of these large
fluctuations in the galaxy density field, self-averaging properties
are well-defined only on scales $r<30$ Mpc/h: in this range we find
scaling properties of the conditional density that decays as a power
law as a function of scale with an exponent around minus one (see
Fig.\ref{Gamma_R1}).  Correspondingly, the PDF of conditional
fluctuations $P(r;N)$ presents a stable shape for scales $r \le 30$
Mpc/h, characterized by a long tail for high $N$ values: this ``fat
tail'' is the signature of the structures present in these
samples. Instead, for $r>30$ Mpc/h, the PDF converges neither to a
Gaussian function nor to a well-defined shape showing clear evidence
of non self-averaging properties of conditional fluctuations.

We interpret this as caused by a systematic effect in that sample
volumes are not large enough for conditional fluctuations, filtered on
such large-scales, to be self-averaging; i.e.,  to contain enough
structures and voids of large size to allow reliable determination of
average (conditional) quantities.  This result implies, for instance,
that the average behaviors of both magnitude and redshift
distributions in the magnitude-limited sample are biased by large
spatial fluctuations and thus that their variance only represent a
lower limit to the real intrinsic variance.  Furthermore we discussed
that K and standard evolutionary corrections \citep{blanton2003} do
not qualitatively affect these behaviors. We pointed out the problems
related to the estimation of amplitude of fluctuations and correlation
properties from statistical quantities which employ the normalization
to the estimation of the sample average. As long as a distribution
inside the given sample is not self-averaging, hence not homogeneous,
the estimation of the two-point correlation function is necessarily
biased by strong finite-size effects. Our results are compatible with
a continuation of the power-law behavior of the conditional density on
scales larger than 30 Mpc/h and incompatible with homogeneity on
scales smaller than $\sim 100$ Mpc/h. Only the availability of larger
samples will allow average correlation properties determined on scales
larger than $\sim 30$ Mpc/h.

Our results, because the imply that galaxy distribution
is inhomogeneous on scales of $\sim 100$ Mpc/h, are perfectly
compatible with a ``Copernican'' principle. Indeed any statistically
stationary inhomogeneous point distribution is compatible with the
principle that there is no special point or direction in the
universe. If for instance there were a ``local hole'' or a particular
large structure around us, this would not imply that the
``Copernican'' principle is violated, but simply that the distribution
is spatially inhomogeneous \citep{pwa}.  These are two different
properties which are sometimes confused in the literature
\citep{ellis,pedro}.

Finally we found that fluctuations in mock galaxy catalogs are
Gaussian for $r>20$ Mpc/h, implying that our results are at odds with
the predictions of the concordance $\Lambda$CDM model of galaxy
formation. This result remains the same when considering redshift
space fluctuations (as for the real data) or real space ones. Indeed
we find that the main difference going from real to redshift-space
occurs for scales smaller than $\sim 5$ Mpc/h, where the exponent of
the conditional density passes from $-1.8$ to about $-1$.

Our results are compatible with a series of analyses of galaxy number
counts in different catalogs, e.g., APM \citep{shanks,maddox}, 2dFGRS
\citep{busswell03}, 2MASS \citep{frith03}, and a sample of galaxies in
the $H$ band \citep{frith06}. In all those surveys count fluctuations
not normalized to the sample average have been considered, and it was
concluded that there are local fluctuations of $\sim 30\%$ extending
over scales of $\sim 200 $ Mpc/h, which are at odds with $\Lambda$CDM
predictions. Furthermore our results are compatible with the results
by \citet{loveday} on the SDSS-DR1 sample, although their
interpretation is different.

Similar persistent spatial fluctuations in the galaxy density field
have been found in the 2dFGRS by \citet{paper_2df_prl,paper_2df_aea}:
this shows that these fluctuations are quite typical of the galaxy
distribution.

In addition the comparison with the model predictions with real galaxy
data, through the analysis of mock galaxy catalogs, which we have
discussed, agrees with that of \citet{einasto1,einasto2,einasto3} who
found, for instance, that super-clusters in real samples extracted
from the 2dFGRS and from the SDSS are more than ten times larger than
in simulated samples constructed from the Millennium simulations. A
similar conclusion was reached by \citet{paper_2df_prl,paper_2df_aea}
on the 2dFGRS.


\section*{Acknowledgments} 
We thank Tibor Antal, Michael Joyce, Andrea Gabrielli, and Luciano
Pietronero for useful remarks and discussions. We are grateful to
Michael Blanton, David Hogg and Martin L\'opez-Corredoira for
interesting comments.  We warmly thank an anonymous referee for a list
of suggestions and criticisms that allowed us to improve the
presentation.  Y.V.B. acknowledges the support by Russian Federation
grants Leading Scientific School --- 1318.2008.2 and RFBR-09-02-00143.
We acknowledge the use of the Sloan Digital Sky Survey data ({\tt
  http://www.sdss.org}), of the NYU Value-Added Galaxy Catalog ({\tt
  http://ssds.physics.nyu.edu/}), and of the Millennium run
semi-analytic galaxy catalog ({\tt
  http://www.mpa-garching.mpg.de/galform/agnpaper/}).

{}


\appendix 

\section{Cosmological corrections}
\label{cosmocorr}

In this appendix we discuss the problem of cosmological corrections to
be applied to the data in some detail. For each galaxy it is observed,
among other quantities, the angular coordinates, the redshift $z$, and
the apparent magnitude $m_r$.  From these data we aim to construct
three-dimensional samples that are not affected by observational
selection effects.  It is observationally established that the galaxy
redshift is linearly proportional to its distance, i.e., the Hubble law
\citep{hubble} $R=c/H_0 z$, where $c$ is the speed of light and $H_0$
is the Hubble constant\footnote{In what follows we denote the Hubble
  constant as $H_0=100 h$ km/sec/Mpc where $h$ is a parameter in the
  range $0.5<h<0.75$ according to observations \citep{hubble2}}. In
the framework of the Friedmann solutions of Einstein field equations,
the linearity of the Hubble law \citep{pee80} is verified only for
very low redshifts. In general, the (metric) distance $R$ depends on
the values of cosmological parameters such as the mass density
$\Omega_m$ and the cosmological constant $\Omega_\Lambda$, so that
$R=R(z; \Omega_m, \Omega_\Lambda)$ \citep[see][]{hoggdistance}.  These
formulas introduce second-order corrections to the linear law that are
generally unimportant at low redshifts, e.g., $z<0.2$, such as the
ones we consider in what follows.

In order to reconstruct the absolute magnitude from the apparent one
we needed to determine the so-called K-correction.  This correction
must be applied because galaxies observed at different redshifts are
sampled, by any particular instrument, at different rest-frame
frequencies. The transformations between observed and rest-frame
broad-band photometric measurements involve terms known as
K-corrections \citep{hms56,hogg_kterm}. In general, if the galaxy
spectrum is known, we can calculate precisely what the K-correction
is. While in the past \citep[see][]{esp} these were known in an
average way, in the case of the SDSS it is possible to reconstruct the
K-correction for each object from the measurements of galaxy
magnitudes in different frequency bands \citep{blanton2005}.

Another correction that has to be considered to determine the absolute
magnitude from the apparent one is related to the way galaxies have
evolved from high to low redshifts \citep{kauffmann,blanton2003}. We
expect that this is a relatively minor problem in our studies because
the maximum redshift we consider is $z=0.2$ and evolutionary
corrections (or E-corrections) are generally believed to be small and
linearly proportional to the redshift. Indeed, as we discuss in what
follows these corrections may only play a role for very bright objects
which can be observed far away from us. There is no well-accepted
model for galaxy evolution and in what follows we adopt the
corrections that are usually used in the literature
\citep[see][]{blanton2003,tegmark2004}. Being applied in an average
way, these corrections have the disadvantage of not taking the galaxy
type into account: spiral, elliptical and irregular galaxies should
have in principle different star-formation histories and thus
different corrections \citep{yoshii}.

It is, however, worth commenting on the derivation of the average
evolution corrections by \citet{blanton2003}. These have been derived
by assuming that the space density is constant (i.e., uniformity), by
including the effect of large-scale fluctuations in some ad-hoc
parameters of a phenomenological behavior of the luminosity-redshift
function and by assuming that unknown evolutionary factors may explain
the residual behaviors that are not taken into account by those
parameters --- see Eq.5 in \citet{blanton2003} and \citet{lin}. Thus
the results for the amount of evolution are based on very strong
assumptions which are reflected in the following: any deviation from
uniformity on a large scale, which is not properly described by the
assumed phenomenological luminosity-redshift function results as a
sign of galaxy evolution; that is, galaxy evolution corrections were
not measured in a way that is free of {\it a priori} assumptions.

Because there is no well-defined way to describe E-corrections, we use
the same type of average functional behavior adopted by other authors
\citep[see][]{tegmark2004} to reach, in the same samples we consider,
conclusions that are substantially different from ours.  We find that
the results for the PDF of conditional fluctuations are basically
unaffected. Although this does not strictly imply that evolution is
not playing any role, this does imply that galaxy fluctuations are not
self-averaging and that galaxy distribution is not uniform in these
samples, at least not in the range of scales that we define properly
below.

Thus the question remains open of whether some more detailed
evolutionary corrections can qualitatively change the results we
get. The basic issue to be considered in this respect is that we
mainly focus the PDF of conditional fluctuations.  While the
E-corrections may change average behaviors as a function of scale, it
is unlikely that they can produce the large amplitude fluctuations of
large spatial extension that we observe. In what follows we present
specific tests computing the effect of average evolution corrections
on the relevant statistical quantities we measure.

\section{Number counts in the magnitude limited sample}
\label{sec:magnlim}

The advantage in using the magnitude limited sample is that one only
considers directly observed quantities, i.e., $m_r, z, \eta, \lambda$,
without K and E corrections that introduce some additional hypotheses
about the shape of galaxy spectrum and the evolution process.  Here we
determine galaxy counts as a function of the apparent magnitude and
the redshift distribution in the ML sample, also determining their
typical fluctuations.  Given the spread in the galaxy luminosity
function, it is not straightforward to derive precise information on
spatial fluctuations and their correlations from these measurements.

\subsection{Magnitude counts}

The analysis of galaxy counts as a function of apparent magnitude
allows us to make an independent estimation, from those based on the
three-dimensional analysis of fluctuations in the survey in VL
samples. However this analysis does not allow us to disentangle
luminosity selection effects from spatial fluctuations. By studying
the variance of counts we are only able to estimate real-space
fluctuations indirectly.

We first divide the angular region of the survey into $N_f=20$
angular subregions of almost equal solid angle. In the $i^{th}$
subregion, of solid angle $\Omega_i$, we compute the differential
counts of galaxies $n_i(m, \Delta m)$ in magnitude bins of size
$\Delta m = 0.25$.  We then compute the average
\be 
\label{avecounts} 
\overline{n(m, \Delta m)} = \frac{1}{N_f} \sum_{i=1}^{N_f}
\frac{n_i(m, \Delta m)}{\Omega_i} \;, 
\ee
and  we estimate  the variance 
\be 
\label{avecounts_Sigma} 
\overline{\Sigma^2(m, \Delta m)} = \frac{1}{N_f-1}
\sum_{i=1}^{N_f} (n_i(m, \Delta m) - \overline{n(m, \Delta m)})^2 \;.
\ee
The normalized variance is 
\be
\label{sigmacounts}
\overline{\sigma^2(m, \Delta m)} = \frac{\overline{\Sigma^2(m,\Delta m)}}
{\overline{n(m, \Delta m)}^2} \;.
\ee
Fig.\ref{ave_counts} shows the average differential number counts of
galaxies as a function of apparent magnitude, which  agrees nicely with
the determination by \citet{strauss2002}.  In particular the
  counts grow as $10^{\alpha m}$ with $\alpha \approx 0.57 \pm 0.01$
  in the magnitude range [14,17].
\begin{figure}
\begin{center}
\includegraphics*[angle=0, width=0.5\textwidth]{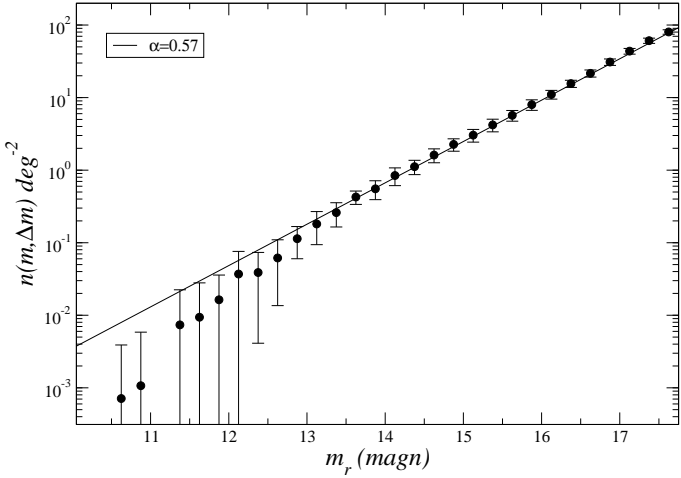}
\end{center}
\caption{Average differential number counts of galaxies as a function
  of apparent magnitude $m_r$ in bins of $\Delta m_r=0.25$ per unit
  solid angle in deg$^2$. The best fit with a behavior of  type
  $10^{\alpha m}$ is shown for $\alpha \approx 0.57 \pm 0.01$.  }
\label{ave_counts}
\end{figure}

In Fig.\ref{var_counts} we present the behavior of $\overline{\sigma(m, \Delta
m)}$ as a function of apparent magnitude.
\begin{figure}
\begin{center}
\includegraphics*[angle=0, width=0.5\textwidth]{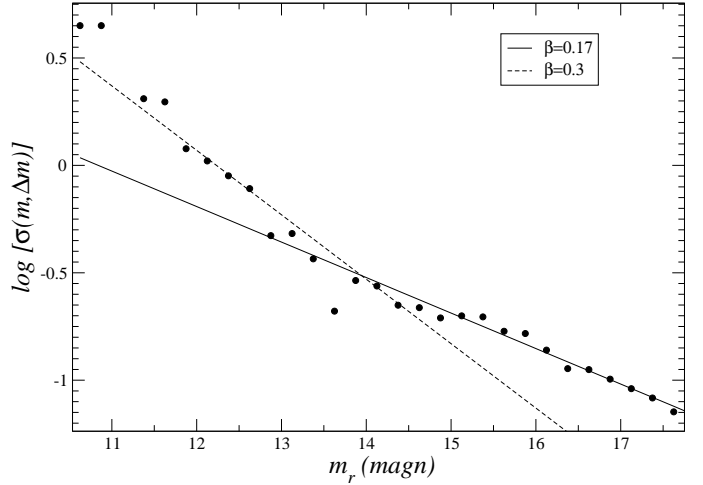}
\end{center}
\caption{Behavior of $\overline{\sigma(m, \Delta m)}$ as a function
  of apparent magnitude (i.e., Eq.\ref{sigmacounts}) in bins of size
  $\Delta m_r=0.25$ per unit solid angle in deg$^2$.}
\label{var_counts}
\end{figure}
The fast decay of $\overline{\sigma(m, \Delta m)}$ at bright
magnitudes (i.e., $m_r < 14$) comes from the dominance of Poisson noise
on the intrinsic variance of the distribution.  It is in fact simple
to show that for a perfectly Poisson distribution of galaxies
(i.e., without any spatial correlation), we get
\be \sigma(m) \sim 10^{-\beta m} 
\ee 
with $\beta =0.3$.  The parameter $\beta$ is in general determined by
the decay of spatial correlations.  For correlated distributions, the
decay is slower than for the Poisson case; i.e., $\beta < 0.3$ but it
is not straightforward to relate the parameter $\beta$ to the exact
value of the correlation exponent in real-space \citep{book}.

One may note from Fig.\ref{var_counts} that, for magnitudes fainter
than $m_r \approx 14$, there are fluctuations of $\sim 15\%$ up to the
faintest magnitude limit of the survey, i.e., $m_r=17.77$: these are
thus persistent up to the deepest scales observed.  This result is not
unexpected, because for many years relatively large fluctuations have
been detected by different authors in many different catalogs. For
instance by studying the POSS-II photographic plates, fluctuations of
$\sim 30 \% $ in the surface galaxy density were observed in the
magnitude range between $16.5-19$ in the $r$ filter \citep{picard91},
although calibration and systematic errors could affect the
photometric determinations from the photographic plates \citep{gd}.

Furthermore a deficiency of bright galaxies around the south galactic
pole was first examined by \citet{shanks} and then by \citet{maddox}
which observed a large deficit in the number counts ($50 \%$ at $B
=16$, $30\%$ at $B =17$) over a 4000 deg$^2$ solid angle.  More
recently in a CCD survey of bright galaxies within the Northern and
Southern strips of the 2dFGRS conclusive evidence was found that there
are fluctuations of about $30\%$ in galaxy counts as a function of
apparent magnitude \citep{busswell03}.  In addition \citet{frith03},
using the bright galaxy counts from the 2 Micron All Sky Survey, found
results indicating a very large `local hole' in the Southern Galactic
Cap (SGC) to $>150$ Mpc/h with a linear size across the sky of $\sim
200$ Mpc/h, suggesting the presence of a potentially huge contiguous
void stretching from south to north, and indicating the possible
presence of significant correlations on scales of the order of $300$
Mpc/h.  Similarly, by studying $H$-band number counts over 0.30
deg$^2$ to $H=19$, as well as $H<14$ counts from 2MASS, concluded that
these counts represent a 4.0 sigma fluctuation implying a local hole
which extends over the entire local galaxy distribution and being at
odds with $\Lambda$CDM predictions \citet{frith06}.  We investigate in
Sec.4, by using the real-space analysis, the relation between these
measurements and fluctuations in real-space, trying to determine
whether the above estimation of the normalized variance is a reliable
statistical measurement of the intrinsic variance of the distribution
or whether there is a systematic effect that may reduce, or enlarge,
the fluctuations measured in this way
\citep[see][]{paper_sdss1,paper_2df_prl,paper_2df_aea}.

It is worth noticing that \citet{Yasuda01}, measured bright galaxy
number counts in two independent stripes of imaging scans along the
celestial equator, one toward the north and the other one toward the
south galactic cap, covering about 230 and 210 square degrees
respectively, from imaging data taken during the commissioning phase
of SDSS. They find that the counts from the two stripes differ by
about 30\% at magnitudes brighter than 15.5. Despite the presence of
these large fluctuations they concluded that the shape of the number
counts-magnitude relation, brighter than $m_r = 16$ is characterized
by the relation expected for a homogeneous galaxy distribution in a
``Euclidean'' universe (for which $\alpha =0.6$) \citep{pee80}. This
result is probably affected by the small number of objects in the
bright end of the counts, which indeed does not exceed a few hundred
galaxies --- see Tables 2 and 6 of \citet{Yasuda01}.  In addition, they
notice that in the magnitude range $16 < m_r < 21$, the galaxy counts
from both stripes agree very well and follow the prediction of the
{\it no-evolution model}, although the data do not exclude a {\it
  small amount of evolution}.  This conclusion  thus contrasts
with the one by \citet{loveday} who, as mentioned,  instead invokes 
a substantial amount of galaxy evolution to explain the radial counts.

Moreover ,it should be noticed that, by measuring the rms scatter of
galaxy number counts in the SDSS-DR1, in different parts of the sky
after correcting for Galactic extinction, \citet{Fukugita} find that
this is consistent with what is expected from the angular
two-point correlation function integrated over circular areas.  They
did not analyze the behavior of the rms scatter as a function of
apparent magnitude, i.e., Eq.\ref{sigmacounts}, and their results show
compatibility of angular correlations with counts fluctuation, but
they do not constraint uniquely spatial correlations. Indeed angular
correlations may be degenerate with respect to three-dimensional
properties \citep[see][]{eckmann,slm}.


\subsection{Redshift distribution}

The analysis of the counts of galaxies as a function of redshift in
the full magnitude-limited survey is a complementary study to the
counts as a function of apparent magnitude. As in the former case, it
is difficult to extract a clear information about correlation
properties of galaxy distribution. However,  analysis of the
redshift distribution of galaxies in different regions on the sky is
an useful instrument for getting a first qualitative information about
the position, sizes and amplitudes of the spatial galaxy number
fluctuations.

For instance, by studying the redshift distribution in the
Durham/UKST Galaxy Redshift Survey, fluctuations were found in the
observed radial density function of  $50 \%$ occurring on
$\sim 50$ Mpc/h scales \citep{Ratcliffe98,busswell03}.  In a similar
way in the 2dFGRS two clear ``holes'' in the galaxy distribution were
detected in the ranges $0.03<z<0.055$, with an under-density of $\sim
40 \%$, and $0.06<z<0.1$ where the density deficiency is about $\sim
25 \%$ \citep{busswell03}.  These two under-densities, detected in
particular in the  2dFGRS southern galactic cap (SGC), are also
clear features in the Durham/UKST survey.  Given that the 2dFGRS SGC
field is entirely contained within the areas of sky observed for the
Durham/UKST survey the similarities in the redshift distributions are
both evidence of the same features in the galaxy distribution.

In Fig.\ref{z_counts_regions} we show the differential number counts,
in bins of  $\Delta z=0.01$ for unit solid angle, as a function of
redshift in the three angular regions R1, R2, and R3.
\begin{figure}
\begin{center}
\includegraphics*[angle=0, width=0.5\textwidth]{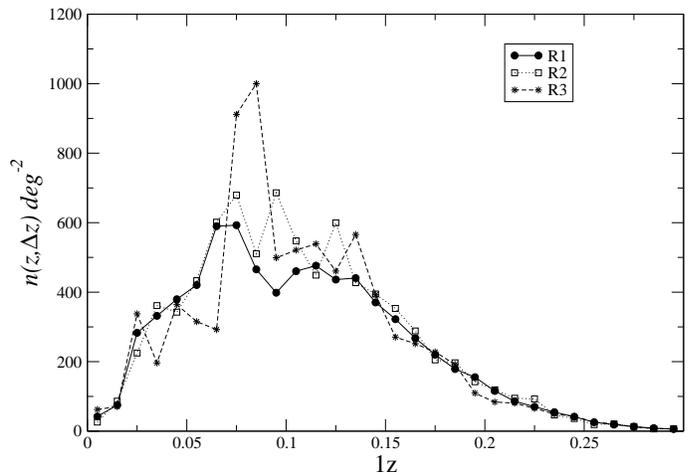}
\end{center}
\caption{Differential number counts as a function of redshift, in bins
  of  $\Delta z=0.01$, for unit solid angle, in the 3 angular
  regions R1, R2 and R3.}
\label{z_counts_regions}
\end{figure}
Although the three angular regions cover different solid angles (in
particular R1 has a solid angle six times larger than R2 and R3), it
is interesting to note that in R3 there is a very large fluctuation
which, as we discuss in Sect.2, corresponds to the famous SDSS Great
Wall \citep{gott}. Other structures of smaller amplitude are visible
in R2 and R3, and we present a more detailed analysis below.  A part
the fluctuations, the behavior of the counts as a function of redshift
involve a convolution with the luminosity selection of the survey.
Thus it generally displays  asymmetric bell-shaped behavior, where
the peak corresponds to the maximum of the luminosity selection of the
survey \citep[see][]{busswell03}.

In Fig.\ref{z_counts_ave} we show the average differential number
counts, in bins of size $\Delta z=0.01$, for unit solid angle. This is
computed similarly to the average counts as a function of apparent
magnitude described above. We divide the angular sky region of the
survey into $N_f=20$ independent and nonoverlapping angular regions
(the $i^{th}$ angular region has solid angle $\Omega_i$).  We then
compute
\be
\overline{n(z, \Delta z)} = \frac{1}{N_f} \sum_{i=1}^{N_f} 
\frac{n_i(z, \Delta z)}{\Omega_i} 
\ee
where $n_i(z, \Delta z)$ represents the counts in the $i^{th}$ sky
region. In Fig.\ref{z_counts_ave} we report the average differential
number counts, in bins of  $\Delta z=0.01$, for unit of solid
angle, as a function of redshift, where  the fluctuations again trace
large-scale structures and the peak at $z\approx 0.07$ corresponds to
the SDSS Great Wall \citep{gott}.

The redshift counts variance is  given by 
\be 
\overline{\Sigma^2(z, \Delta z)} = \frac{1}{N_f-1}
\sum_{i=1}^{N_f} (n_i(z, \Delta z) - \overline{n(z, \Delta z)})^2 \;.
\ee
The normalized variance  is thus 
\be
\overline{\sigma^2(z, \Delta z)} = \frac{{\overline{\Sigma^2(z,\Delta z)}}}
{\overline{n(z, \Delta z)}^2 } \;.
\ee
In general the variance for a point distribution is the sum of the
intrinsic variance due to correlations and to Poisson noise. Here we
subtract the Poisson term, so  we only consider  the intrinsic
variance due to correlations.  In Fig.\ref{z_counts_var} we present
the normalized (intrinsic) standard deviation for different choices of
$\Delta z$.  When the redshift bin is increased to $\Delta z =0.05$
(which corresponds to $\Delta R \approx 150$ Mpc/h) fluctuations are
still of  $\sim 15 \%$, and they  persist at the
different scales sampled by the survey, in agreement with the results
obtained by the apparent magnitude counts analysis and with the
analysis in other galaxy redshift surveys
\citep{Ratcliffe98,busswell03,paper_2df_aea}.
\begin{figure}
\begin{center}
\includegraphics*[angle=0, width=0.5\textwidth]{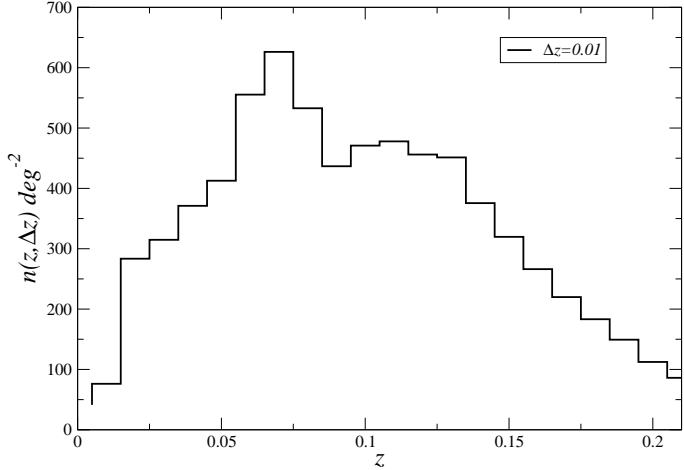}
\end{center}
\caption{Average differential number counts, in redshift bins of size
  $\Delta z=0.01$, for unit of solid angle, as a function of
  redshift.}
\label{z_counts_ave}
\end{figure}
\begin{figure}
\begin{center}
\includegraphics*[angle=0, width=0.5\textwidth]{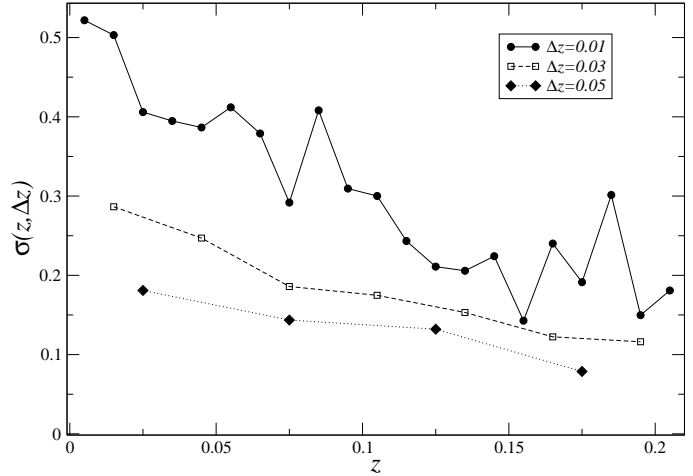}
\end{center}
\caption{Standard deviation of the differential number counts, in
  redshift bins of size $\Delta z=0.01, 0.03, 0.05$, for unit of solid
  angle, as a function of redshift. Poisson noise has been subtracted
  so this only contains the contribution due to galaxy correlations. }
\label{z_counts_var}
\end{figure}


\section{The luminosity function}
\label{sec:lumfun}

One of the main problem in the study of galaxy structures is to
disentangle spatial properties of galaxies from their luminosity
distribution.  Thus an important quantity to be determined is the
galaxy luminosity function $\phi(L)$ and the quantity $\phi(L)dL$
provides the probability that a galaxy has luminosity $L$ in the range
$dL$. In general an assumption is made that the ensemble average
number of galaxies for unit volume and unit luminosity can be written
as
\be 
\label{approxnu}
\langle \nu(\vec{r},L) \rangle = \langle n( \vec{r} ) \rangle 
\times \langle \phi(L)\rangle \;,
\ee 
where $\langle n(\vec{r}) \rangle$ is the ensemble average density and
$\langle \phi(L)\rangle$  the ensemble average luminosity function.
This implies the independence between space and luminosity
distributions, i.e., that galaxy positions are independent of their 
luminosities. Although there is  clear evidence of  a 
correlation between them  (as for instance the brightest elliptical
galaxies are found in the center of rich galaxy clusters) it has been
tested that this is nevertheless a reasonable assumption in the galaxy
catalogs available so far \citep[see][]{book}.  To go beyond this assumption
one should use the multi-fractal formalism as in \citet{slp96}. 

An additional, much stronger, assumption often adopted is that the
space density is a constant, i.e.,  $\langle n( \vec{r} ) \rangle =
const$. This assumption is for instance at the basis of the so-called
standard minimum variance estimator
\citep{davishuchra,blanton2003,loveday}.  It is clear that we want to
avoid making this further assumption because we want to test whether
the space density is (or can be approximated by) a simple constant. It
is also evident that if this assumption is inconsistent with the
sample data properties, all results derived from methods encoding it
are intrinsically biased.

To determine the shape of the luminosity function, the so-called
inhomogeneity-independent method is commonly employed
\citep{loveday,blanton2003} which uses a modified version of
Eq.\ref{approxnu}, namely that
\be
\label{approxnu2}
\nu(R,L) = n(R) \times \phi(L) \;.
\ee 
where $n(R)$ is the density as function of the radial (metric)
distance $R$ and $\phi(L)$ the luminosity function. This can be a
useful working hypothesis.  Under this approximation, in a VL sample
the luminosity function can be written as
\be
\label{eq:phi}
\phi^{VL}(L) =
\frac{\phi(L) \int_{R_{min}}^{R_{max}} n(R) \Omega R^2 dR}{N}
\ee
where $N$ is the total number of galaxies in the VL sample and
$\Omega$ its solid angle. In this way, even when $n(R)$ is highly
fluctuating, one may recover the shape of $\phi(L)$ as spatial
inhomogeneities cancel out in the ratio given in Eq. \ref{eq:phi}.
Thus by making the normalized histogram of
the number of galaxies in luminosity bins in each VL sample, we get
$\phi^{VL}(L)$. Then we look for the best fit in all the VL samples
with the Schechter function \citep{sch76}
\be
\phi(L) = A \times L^{\alpha} \exp(-L/L^*) \;.
\ee
For this determination we used other VL than those listed in
Table~\ref{tbl_VLSamplesProperties1}; namely, we constructed VL samples
each with only one magnitude in range.  We then find (see
Fig.\ref{LF_SDSS})  in the K-corrected catalog with no E-corrections
that the best-fit parameters are $\alpha=1.22 \pm 0.02$ and
$M^*=-20.63 \pm 0.02$,in good agreement with previous determinations
\citep[see][]{loveday}.
\begin{figure}
\begin{center}
\includegraphics*[angle=0, width=0.5\textwidth]{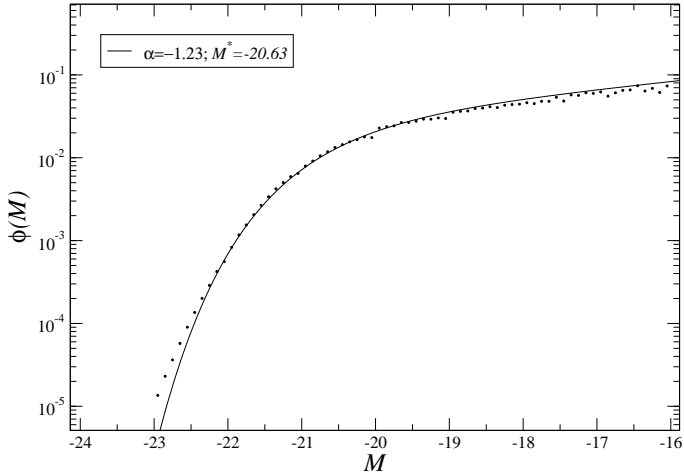}
\end{center}
\caption{Luminosity function in the SDSS K-corrected catalog and its
  best-fit estimation with a Schecther function.}
\label{LF_SDSS}
\end{figure}

To conclude this discussion we note that while the effect of
inhomogeneities is fairly taken into account in Eqs.\ref{approxnu}-
\ref{approxnu2}, the amplitude of the luminosity function is usually
estimated under the assumption that this is a constant proportional to
the average density. We have seen that this situation cannot be 
satisfied in the data; i.e., when $n(\vec{r})$ has a clear scale
dependence, the amplitude of the luminosity function gives a
systematically biased estimation of the average density.

\end{document}